\documentclass[prl,aps,amssymb,twocolumn,superscriptaddress, notitlepage, longbibliography]{revtex4-2}

\usepackage{bm,amssymb,amsmath,dsfont,mathrsfs,amstext,latexsym,physics,relsize}
\usepackage{graphicx}

\usepackage[usenames,dvipsnames]{xcolor}
\usepackage[colorlinks=true,citecolor=teal,urlcolor=teal,linkcolor=teal]{hyperref}

\let\Re\relax
\let\Im\relax
\DeclareMathOperator{\Re}{Re}
\DeclareMathOperator{\Im}{Im}
\usepackage{orcidlink}

\newcommand{\figpanel}[2]{\hyperref[#1]{\ref{#1}#2}}

\let\oldsection\section
\renewcommand{\section}[1]{\oldsection{\MakeUppercase{#1}}}
\let\oldsubsection\subsection
\renewcommand{\subsection}[1]{\oldsubsection{\MakeUppercase{#1}}}

\begin{document}

\title{Tailoring Quantum Chaos With Continuous Quantum Measurements}

\author{Preethi Gopalakrishnan\orcidlink{0009-0000-4053-235X}}
\email{preethi.gopalakrishnan@uni.lu}

\author{Andr\'as Grabarits\orcidlink{0000-0002-0633-7195}}
\email{andras.grabarits@uni.lu}
\affiliation{Department  of  Physics  and  Materials  Science,  University  of  Luxembourg,  L-1511  Luxembourg, Luxembourg}

\author{Adolfo~del Campo\orcidlink{0000-0003-2219-2851}} 
\email{adolfo.delcampo@uni.lu}
\affiliation{Department  of  Physics  and  Materials  Science,  University  of  Luxembourg,  L-1511  Luxembourg, Luxembourg}
\affiliation{Donostia International Physics Center,  E-20018 San Sebasti\'an, Spain}

\begin{abstract}
We investigate the role of quantum monitoring in the dynamical manifestations of Hamiltonian quantum chaos. Specifically, we analyze the generalized spectral form factor, defined as the survival probability of a coherent Gibbs state under continuous energy measurements. We show that quantum monitoring can tailor the signatures of quantum chaos in the dynamics, such as the extension of the ramp in the spectral form factor, by varying the measurement strength and detection efficiency. In particular, a typical quantum trajectory obtained by monitoring with unit efficiency exhibits enhanced quantum chaos relative to the average dynamics and to unitary evolution without measurements. Our results hold for systems across symmetry classes and for other diagnostics of quantum chaos, such as out-of-time-order correlators and Liouvillian spectral statistics.
\end{abstract}

\maketitle
Quantum chaos manifests itself in the spectral statistics of isolated quantum systems. In particular, chaotic systems exhibit level repulsion, leading to a Wigner-Dyson distribution of nearest-level spacings, in contrast to the Poissonian statistics characteristic of integrable systems \cite{Mehta_1991_Random_Matrices_Review,Haake}. This connection between chaos and spectral statistics has motivated the development of diagnostic tools based on the Fourier transform of energy spectra \cite{Leviandier86,WilkieBrumer91,Alhassid1992,Ma95,Prange97}. In this context, a prominent tool is the Spectral Form Factor (SFF), which captures correlations in the energy spectrum and exhibits a characteristic dip-ramp-plateau in chaotic quantum systems \cite{Haake,Cotler2017BH}.

Signatures of quantum chaos also appear in quantum dynamics beyond static spectra. While they can be appreciated in the time-evolution of quantum observables, they are often more pronounced in quantities such as the survival probability and the Loschmidt echo \cite{Ma95,delcampo17,TorresHerrera20,Adway25}, the characteristic function of the work statistics \cite{Chenu17work,Chenu19} and frame potentials \cite{Cotler2017chaos}. Indeed, the SFF can be interpreted as the fidelity between a coherent Gibbs state and its time-evolved counterpart \cite{Xu21SFF}, linking spectral statistics directly to dynamical behavior and motivating experimental studies across several platforms \cite{Vasilyev20,Joshi22,Dong25}.

Realistic quantum systems, however, are rarely isolated, as they are generally embedded in an environment \cite{BP02,Zurek03}. This raises a natural question: how do the signatures of quantum chaos persist under open-system dynamics? Early studies revealed that classical chaos can emerge from quantum systems due to decoherence  \cite{Karkuszewski02,Habib06,Lewalle18}, while other works explored how chaos influences decoherence itself \cite{Cucchietti03,Xu19,delCampo2020}. 

Two complementary approaches have been developed to study spectral statistics in open quantum systems. One generalizes spectral statistics to the generator of evolution in open quantum systems \cite{Can19,Can19b,Sa19,SaProsen20,SaProsen21}, while the other examines how Hamiltonian quantum chaos are modified under non-unitary dynamics \cite{Xu21SFF,CaoXu22,Cornelius21,MatsoukasRoubeas23,Matsoukas23SFFSA,Zhou2024}. Here we adopt the latter approach, focusing on dynamical signatures of spectral statistics \cite{MatsoukasRoubeas24PQC}.

The signatures of quantum chaos in an isolated system are generally altered in an open setting. The interplay between chaos and decoherence can be probed via the SFF for open systems, which involves the Ulhmann fidelity between an initial coherent Gibbs state and its time evolution, generally described by a mixed state \cite{Xu21SFF}. Generic dephasing mechanisms have been shown to quickly suppress chaotic signatures, but when dephasing occurs on the energy eigenbasis, a smooth crossover emerges: the depth of the correlation hole decreases, and the onset of the ramp is delayed \cite{Xu21SFF}. These effects are robust and persist in related tools for diagnosing quantum chaos, such as the local energy distribution in multipartite quantum systems \cite{CaoXu22}.

The question arises: What kind of dynamics can enhance signatures of quantum chaos? Earlier works have shown that non-Hermitian evolution, associated with energy dephasing in the absence of quantum jumps, can amplify chaotic signatures \cite{Cornelius21,MatsoukasRoubeas23}. Such evolution effectively implements in the laboratory the kind of spectral filters used in computational many-body physics \cite{Prange97,Gharibyan18,Matsoukas23SFFSA,VallejoFabila24,Orman2025}. However, diagnosing chaos with the SFF requires long-time dynamics, up to the time associated with the onset of the plateau, set by the inverse of the average level spacing. Non-Hermitian evolution can enhance chaos, but it relies on post-selecting trajectories with no quantum jumps, whose probability decays exponentially with time, making this approach experimentally unrealistic.

Motivated by the recent experimental progress in studying the SFF \cite{Vasilyev20,Joshi22,Dong25}, in this Letter, we investigate the fate of quantum chaos under continuous energy measurement and demonstrate that the signatures of chaos in the SFF are enhanced in a typical quantum trajectory. As a result, quantum monitoring provides a realistic, experimentally feasible route to amplify quantum chaos. Furthermore, we show that the degree of chaotic behavior exhibited in the dynamics of a quantum system can be controlled by varying the measurement strength and efficiency.

{\it Evolution under continuous quantum measurements.---} The evolution of a quantum system under continuous measurements of a set of observables $\{ A_\alpha \} $ is described by the stochastic master equation (SME) \cite{Jacobs2014}
\begin{eqnarray}\label{Eq:SME}
\mathrm{d}\rho(t) = \mathcal{L}[\rho(t)]\, \mathrm{d}t + \mathcal{I}[\rho(t)]\, \mathrm{d}W_t,
\end{eqnarray}
where the Liouvillian $\mathcal{L}[\cdot]$ accounts for the linear open dynamics
    $\mathcal{L}[\rho(t)] = -i [H, \rho(t)] - \sum_\alpha \gamma [A_\alpha, [A_\alpha, \rho(t)]]$, and the innovation term $I[\cdot]$ describes the nonlinear stochastic backaction due to measurement: $I [\rho(t)] = \sum_\alpha \sqrt{2\gamma} \left( \{A_\alpha, \rho(t)\} - 2 \text{Tr}[A_\alpha \rho(t)] \rho(t) \right)$.
Here, $ \mathrm{d}W_t $ denotes the infinitesimal increment of the Wiener process $W_t$, which satisfies the Itô correlation, $\mathbb{E}\left[ \mathrm{d}W_t\right] = 0$ and $\mathbb{E}\left[ \mathrm{d}W_t^2\right] =  \mathrm{d}t$, such that
    $\mathbb{E} \left[ \mathrm{d}W_t \, \mathrm{d}W_s \right] = \delta_{t,s} \, \mathrm{d}t.$  The time evolution of $\rho(t)$ under a single realization of the measurement noise $\mathrm{d}W_{t}$ defines a quantum trajectory. Each trajectory corresponds to one possible stochastic evolution of the system conditioned on the measurement outcomes, as described by the SME in Eq. (\ref{Eq:SME}). Averaging over independent trajectories is equivalent to disregarding the measurement records, since $\mathbb{E}[{\mathrm d}W_t] = 0$,  and yields an evolution for the ensemble-averaged state according to the standard Lindblad master equation $\mathrm{d}\rho(t) = \mathcal{L}[\rho(t)]\, \mathrm{d}t$ \cite{Wiseman2009}. 

We focus on monitoring a single observable, the energy of the system. Thus, $\{A_\alpha\} = H $ and the evolution of a single trajectory is given by the SME 
    \begin{eqnarray}\label{eq:SME_nonlin}
 \mathrm{d}\rho(t) =&& -i[H,\rho(t)] \mathrm{d}t - \gamma[H,[H,\rho(t)]] \mathrm{d}t\nonumber\\
&&+ \sqrt{2\gamma} \left[\{H, \rho(t)\}-2\langle H\rangle\rho(t)\right] \mathrm{d}W_t.
\end{eqnarray} 
As shown in \cite{SM}, this evolution is equivalent to the dynamics that arises in the energy-driven collapse model for state reduction \cite{Gisin84,Milburn91,Percival94,Diosi98,Brody02,Adler03}. 
Consider a system described by the a Hamiltonian $H$ with spectral decomposition  $H = \sum_{n=1}^{d} E_n \lvert n\rangle \langle n\rvert$, where $d$ is the dimension of the system Hilbert space $\mathcal{H}$, $\{E_n\}$ denote the energy eigenvalues and $\{\lvert n\rangle\}$ the corresponding eigenstates. As detailed in \cite{SM}, for a generic initial quantum state $\rho(0) = \sum_{nm} \rho_{nm}(0)\lvert n\rangle \langle n\rvert,$ its evolution under continuous energy measurements is given by
\begin{widetext}
\begin{equation}\label{Eq:Rho_t}
    \rho(t) = \frac{ \sum_{nm}\rho_{nm}(0)e^{-i(E_n -E_m)t - 2\gamma t(E_n^2 + E_m^2) + \sqrt{2\gamma}W_t(E_n+E_m)}}{ \sum_k \rho_{kk}(0)e^{-4\gamma tE_k^2 + 2\sqrt{2\gamma}W_tE_k}}\lvert n\rangle \langle m\rvert.
\end{equation}
\end{widetext}
where $W_t$ is the stochastic integral, $W_t = \int_0^t\mathrm{d}W_{t'}$. 
Note that the evolution in Eq.~(\ref{Eq:Rho_t}) remains pure along each trajectory, since for a given realization of $W_t$, $\rho(t)$ can be written as a normalized projector, $\rho(t) = |\psi_t\rangle\langle \psi_t|/\langle \psi_t|\psi_t\rangle$. Stochasticity affects only the amplitudes in $|\psi_t\rangle$, and purity is lost only upon averaging over many noise realizations, resulting in the mixed ensemble-averaged state described by the Lindblad equation $ \mathrm{d}\rho(t) = -i[H,\rho(t)] \mathrm{d}t - \gamma[H,[H,\rho(t)]] \mathrm{d}t$. The latter also provides an ensemble description when the quantum evolution is timed by a realistic clock subject to errors \cite{Egusquiza99,Egusquiza03,Xu19}. 

\emph{Spectral Form Factor under continuous measurements.---} To probe signatures of quantum chaos under stochastic dynamics, we employ the fidelity-based definition of the spectral form factor (SFF), which remains meaningful for arbitrary quantum dynamics and is therefore directly applicable to continuous measurements. A natural generalization of the SFF to open quantum systems is given by the fidelity between the initial coherent Gibbs state, $\lvert \psi_\beta \rangle = \sum_n \frac{e^{-\beta E_n/2}}{\sqrt{Z(\beta)}} \lvert n\rangle$ with $Z(\beta) = \mathrm{Tr}[e^{-\beta H}]$,  
and its time-evolved state under stochastic dynamics. The corresponding fidelity,
$F_\beta(t,W_t) = \langle \psi_\beta \rvert \rho(t) \lvert \psi_\beta \rangle$, is then given by
\begin{equation}\label{Eq:SFF}
    F_\beta(t,W_t) = \frac{ \left|\sum_{n} e^{-(\beta+ it - \sqrt{2\gamma}W_t)E_n}e^{-2\gamma tE^2_n}\right|^2}{Z(\beta)Z(\beta-\sqrt{8\gamma}W_t,\gamma)}, 
\end{equation}
where we define the trajectory-dependent ``dephased partition function'' $Z(\beta-\sqrt{8\gamma}W_t,\gamma)=\sum_ne^{-(\beta -\sqrt{8\gamma}W_t)E_n-4\gamma E^2_nt}$. 

\begin{figure*}
    \centering
    \includegraphics[width=2\columnwidth]{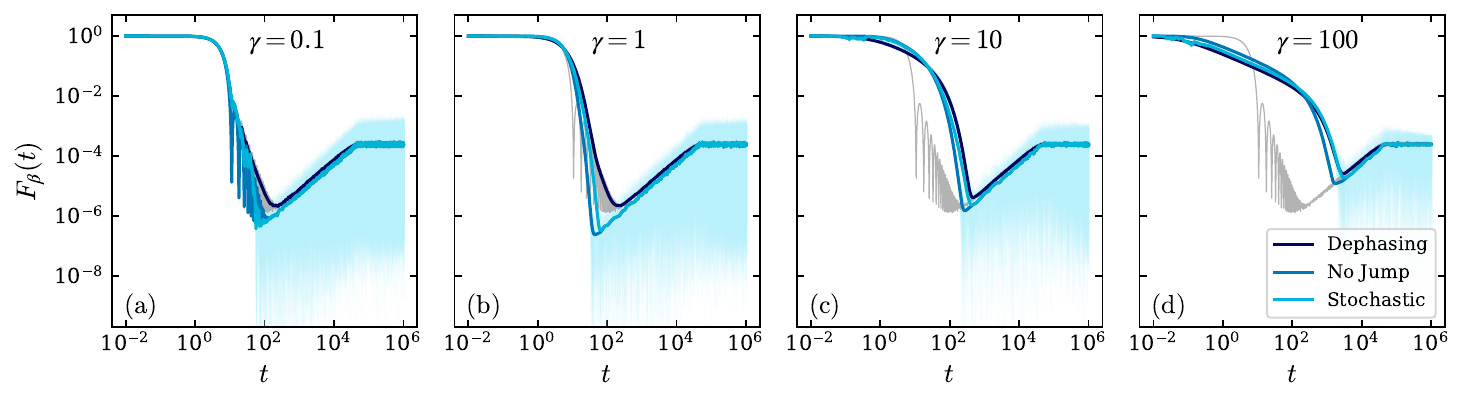}
    \caption{Enhancement of quantum chaos under continuous monitoring in an energy-dephasing process. The time evolution of the spectral form factor (SFF) is shown for different measurement strengths $\gamma$ in the SYK model with $N=26$ and $\beta=0$, averaged over 250 Hamiltonian ensembles and noise realizations. The light colored lines in the background depict various individual stochastic trajectories. The SFF for unitary evolution in a closed system ($\gamma = 0$) is shown by the gray curve.}
    \label{fig:SFF_comparison}
\end{figure*}

Let us next discuss the main features of the SFF under continuous monitoring. For an isolated system ($\gamma = 0$), the SFF displays a characteristic dip-ramp-plateau structure \cite{Haake,Cotler17}, in which the extension of the ramp, from the dip time $t_\mathrm{d}$ to the plateau time $t_\mathrm{p}$, can be used to quantify quantum chaos~\cite{Suntajs20}. 
Energy dephasing is known to gradually suppress the ramp in the SFF, making it shallower and delaying its onset, without affecting the plateau \cite{Xu19,Xu21SFF}. In the context of continuous quantum measurements, this evolution arises from disregarding all measurement outcomes. 

Another reference evolution corresponds to null-measurement conditioning, in which the evolution is deterministic, nonlinear, and non-Hermitian \cite{Carmichael09}. It describes quantum trajectories under the monitoring of energy with no quantum jumps and is thus restricted to short-time dynamics. Such evolution effectively filters the contribution to the SFF of each eigenvalue with a time-dependent Gaussian factor, which, for certain values of $\gamma$, prolongs the extension of the ramp, thus enhancing the manifestations of quantum chaos in the quantum evolution \cite{Cornelius21,MatsoukasRoubeas23}. The corresponding SFF can be obtained from Eq.~(\ref{Eq:SFF}) by setting $W_t=0$. Jump-free trajectories are, however, rare as they are exponentially suppressed as a function of time. Their use to enhance quantum chaos is at odds with the need to probe long time scales in the SFF, given that $t_p$ generally scales with $d$ and thus grows exponentially with system size. 

The effect of monitoring quantum chaos in the SFF, leading to Eq.~(\ref{Eq:SFF}), can be understood as supplementing the monitored dynamics under null-measurement conditioning with stochastic temporal fluctuations in the equilibrium inverse temperature. 
As we next illustrate, a typical trajectory under continuous energy measurements, for a suitable measurement strength, extends the ramp duration in the SFF. Thus, quantum monitoring provides a natural framework to enhance the manifestations of quantum chaos probed by the SFF, without the restriction to short-time evolution intrinsic to null-measurement conditioning.  

As a test bed, we consider the maximally chaotic Sachdev-Ye-Kitaev (SYK) model \cite{Chowdhury22},
\begin{equation}
H = \sum_{1\leq k < l < m < n \leq N} J_{klmn} \, \chi_k \chi_l \chi_m \chi_n,
\end{equation}
which describes $N$ Majorana fermions $\{\chi_i\}$ coupled through random all-to-all quartic interactions. The couplings $J_{klmn}$ are independently drawn from a Gaussian distribution with zero mean, $\overline{J_{klmn}}=0$, and variance $\overline{J_{klmn}^2} = 3! J^2/N^3$, where we set $J=1$ for convenience. The isolated SYK model exhibits maximal quantum chaos, displaying random-matrix spectral statistics and a Lyapunov exponent that saturates the universal bound on chaos \cite{MSS16,Tsuji18}. Its realization has been proposed in ultracold atoms \cite{Danshita17}, digital quantum simulators \cite{GarciaALvarez17}, and disordered graphene flakes \cite{Kruchkov23}. Experimental progress has been reported using NMR \cite{Luo2019} and superconducting qubits \cite{Asaduzzaman24}, which provide a versatile platform for exploring the interplay between continuous measurement and many-body quantum chaos. 

Figure~\ref{fig:SFF_comparison} illustrates the impact of continuous monitoring on the SFF in the SYK model. We compare the SFF under stochastic evolution with two reference dynamics: (a) no-jump evolution, described by the non-Hermitian dynamics $\frac{\mathrm{d}}{\mathrm{d}t}\rho = -i(H_{\rm eff}\rho - \rho H_{\rm eff}^\dagger) + i\mathrm{Tr}[(H_{\rm eff} - H^\dagger_{\rm eff})\rho]\rho,$ where $H_{\rm eff} = H - i\gamma H^2$, and  (b) the dissipative Lindblad evolution governed by the master equation, $ \frac{\mathrm{d}}{\mathrm{d}t}\rho = -i[H,\rho] - \gamma[H,[H,\rho]]$. Note that $F_\beta(t)$ involves a double average of the SFF, one over the  Hamiltonian ensemble due to the disordered couplings  $J_{klmn}$ and the second over stochastic trajectories.

For weak measurements, $\gamma\ll1$, the stochastic trajectories closely follow the evolution with no jumps as shown in Fig.~\ref{fig:SFF_comparison}, exhibiting the characteristic dip-ramp-plateau structure associated with chaotic dynamics. This agreement indicates that in this regime, quantum jumps are rare, and the dynamics is effectively governed by the non-Hermitian evolution in the absence of jumps.  This yields a fast decay of the SFF towards the dip, suppressing the nonuniversal part of the SFF associated with the Fourier transform of the average local energy density; see \cite{SM}. As a result, the ramp is prolonged as the dip time occurs earlier. By contrast, the onset of the plateau is unaffected.

Figure~\ref{fig:SFF_comparison} also shows that stochastic evolution exhibits a crossover at $\gamma\approx 1$ with increasing measurement strength $\gamma$. This is quantified in Fig.~\ref{Fig:Diptime}(a) by the ratio $t_\mathrm{d}/t_\mathrm{p}$ of the dip and plateau times. As $\gamma$ increases, quantum jumps become more frequent, leading to stochastic deviations from the no-jump limit. For $\gamma\leq1$, the measurement-induced dynamics enhances the signatures of quantum chaos beyond the ensemble average governed by energy dephasing. Remarkably, the ramp duration is also enhanced relative to the unitary evolution shown in gray. 
For larger $\gamma$, the ramp is reduced relative to the measurement-free case, and the SFF behaves similarly in the three cases involving measurements, as shown in Fig.~\ref{fig:SFF_comparison}(c)-(d). These features are not specific to the SYK model and persist in other classes of chaotic models; see \cite{SM}.

We note that the difference between the doubly averaged $F_\beta(t,W_t)$ and the dephasing Lindblad evolution is due to the breakdown of the annealed approximation with respect to the Wiener process \cite{SM}. With this approximation, the average of Eq.~(\ref{Eq:SFF}) reduces the evolution exactly to that under energy dephasing. However, the fact that the annealed approximation breaks down at all times enhances the signatures of quantum chaos beyond the ensemble average.

\begin{figure}
    \centering
    \includegraphics[width=1\columnwidth]{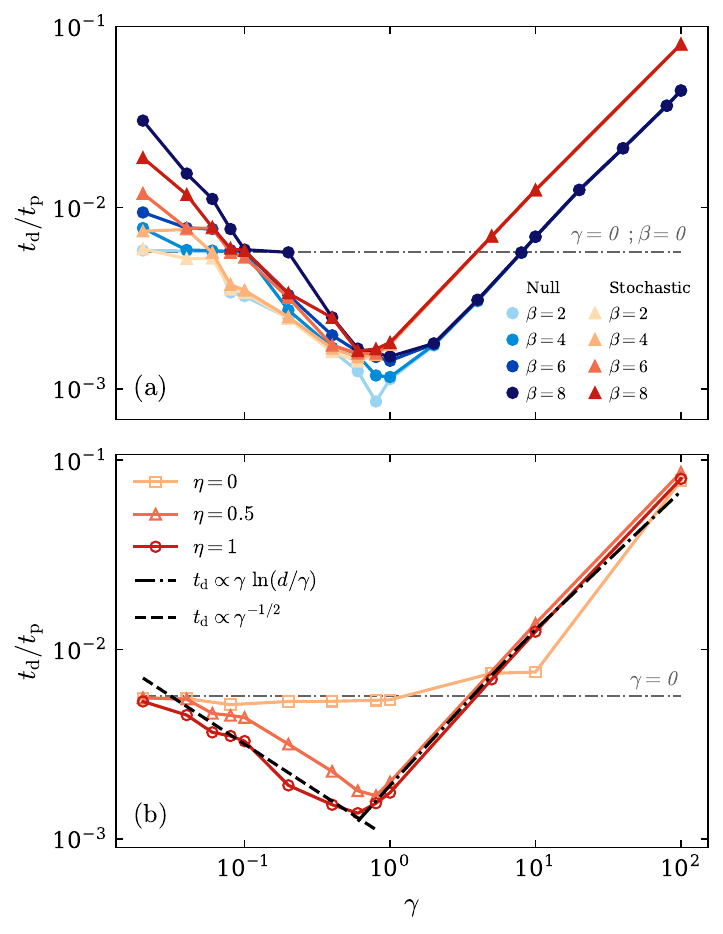}
    \caption{(a) Tailoring quantum chaos by tuning the measurement strength. The ramp duration in the SFF is characterized by the ratio $t_\mathrm{d}/t_\mathrm{p}$  of the plateau and dip times in the monitored SYK model, averaged over 250 stochastic trajectories. It exhibits a non-monotonic behavior as a function of the measurement strength $\gamma$, reaching a maximum for $\gamma_{\rm opt}\sim\mathcal{O}(1)$, when $t_\mathrm{d}/t_\mathrm{p}$ is minimized.  The dependence for small $\gamma$ on the inverse temperature is suppressed for values $\gamma>\gamma_{\rm opt}$. The null measurement limit is shown in blue. (b) Ratio $t_\mathrm{d}/t_\mathrm{p}$  as a function of the measurement strength $\gamma$ for different values of the measurement efficiency $\eta$. In the absence of measurement backaction ($\eta = 0$), the dip time remains nearly constant in the weak measurement regime. By contrast, for finite efficiency ($\eta >0$), $t_\mathrm{d}$ exhibits a pronounced nonmonotonic dependence on $\gamma$. The unitary value of $t_\mathrm{d}/t_\mathrm{p}$ at $\beta=0$ is shown by the gray line.}
    \label{Fig:Diptime}
\end{figure}

The crossover in the SFF under stochastic monitoring arises from the competition between measurement-induced energy filtering at the trajectory level and energy-basis dephasing emerging upon stochastic averaging. The pronounced manifestation of quantum chaos originates from the measurement-induced Gaussian filter $e^{-2\gamma t E_n^2}$ acting on each individual trajectory, with or without jumps; see Eqs.~\eqref{Eq:Rho_t} and \eqref{Eq:SFF}. This provides a physical realization of the spectral filtering commonly used in numerical analysis \cite{Prange97,Gharibyan18,Matsoukas23SFFSA,VallejoFabila24,Orman2025}. Although stochastic averaging ultimately produces dephasing of the form $e^{-\gamma t(E_n-E_m)^2}$, its observable impact depends on how the associated decoherence time $t_{\rm dec}=1/(4\gamma {\rm Var}_\beta H)$ \cite{SM} compares with the intrinsic spectral timescale of the SFF, the dip time, $t_\mathrm{d}$.

When monitoring is weak ($t_\mathrm{dec} \gg t_{\rm d}$), dephasing remains negligible over the timescale relevant for the dip–ramp structure. The dynamics is then dominated by the non-Hermitian damping $e^{-2\gamma t(E_n^2 + E_m^2)}$, which depends on absolute values of the energies rather than energy differences. This acts as an effective spectral filter that suppresses contributions from the edges of the spectrum while largely preserving coherent interference between nearby levels, accelerating the decay of the nonuniversal SFF component, associated with the average density of states, and enhancing the ramp.

As the monitoring strength increases such that $t_\mathrm{d}\gtrsim t_{\rm dec}$, stochastic averaging produces genuine energy-basis dephasing that progressively suppresses off-diagonal spectral coherences. In this regime, the filtering effect is overtaken by decoherence, and the SFF gradually approaches the behavior expected under energy dephasing, with reduced spectral correlations and a shortened ramp. The crossover observed around $\gamma \sim O(1)$ thus reflects the transition from a filtering-dominated regime, where monitoring enhances the signatures of quantum chaos, to a decoherence-dominated regime in which spectral correlations are increasingly washed out; see \cite{SM}. \\


\emph{Inefficient Measurements.---}
We next examine continuous measurements with finite detection efficiency $\eta\in[0,1]$, which provides a natural means to tune the measurement backaction and the manifestation of quantum chaos. For any $\eta$, the system's evolution is governed by the master equation,
${\rm d}\rho = -i[H,\rho]{\rm d}t - \gamma[H,[H,\rho]]{\rm d}t + \sqrt{2\eta\gamma}\left(H\rho + \rho H - 2\langle H\rangle_t \rho\right){\rm d}W_t$,
where $\eta$ represents the fraction of measurement outcomes registered in the detector. This yields the SFF 
\begin{equation}
F_\beta(t,W_t,\eta)=
\frac{\sum_{n,m} e^{-\beta (E_n+ E_m)}\, K_t(E_n,E_m)}
{ Z(\beta)\sum_{n} e^{-\beta E_n}\, K_t(E_n,E_n)},
\end{equation}
where $
K_t(E_n,E_m):=\exp\!\Big[-i(E_n- E_m)t-\gamma t\, (E_n-E_m)^2 -\gamma\eta t \,(E_n+ E_m)^2 +\sqrt{2\gamma\eta}\,(E_n+ E_m)\,W_t
\Big].$ 
Varying $\eta$ allows one to continuously tune the influence of the measurement backaction. 
For $\eta=0$, the dynamics reduce to purely dephasing Lindblad evolution, which minimizes quantum chaos. 
At unit efficiency ($\eta=1$), a quantum trajectory is described by a pure state, and quantum monitoring most strongly reveals the signatures of chaos.
This transition, shown in Fig.~\ref{Fig:Diptime}(b), highlights the role of $\eta$ as a tunable parameter governing the onset of chaos in a continuously monitored quantum system. For finite $\eta$ and weak dissipation, the effective filtering leads to the dip time scaling $t_{\mathrm{d}} \sim \gamma^{-1/2}$. The crossover in which signatures of quantum chaos are maximally pronounced occurs at $\eta\approx1$. For larger $\gamma$, the evolution is governed by dephasing and the dip time scales as 
$t_{\mathrm{d}} \sim \gamma~\mathrm{ln}(d/\gamma)$, where $d$ is the Hilbert space dimension of the Hamiltonian; see \cite{SM}. This regime reflects a Zeno-like suppression of chaos signatures, in which strong monitoring delays the onset of universal spectral correlations.

In conclusion, we have investigated the continuous monitoring of a chaotic quantum system and identified the agency of the observer in tailoring quantum chaos. 

The amplitude of the ramp in the SFF, a manifestation of level repulsion, can be controlled by the monitoring agent by varying the strength and efficiency of the continuous energy measurement. Remarkably, a typical quantum trajectory leads to an enhancement of quantum chaos signatures not only after the disorder average but also when compared to the unitary Hamiltonian dynamics in the absence of measurements. Our conclusions hold for other probes of quantum chaos including  out-of-time-order correlators and Liouvillian spectral statistics, as shown in the End Matter. Our results show how signatures of quantum chaos in the dynamics can be controlled by an external observer, and should find applications in foundations of physics and statistical mechanics \cite{Haake,DAlessio16,Zurek25}, black hole physics \cite{Cotler17,Chowdhury22}, the study of complexity in quantum systems \cite{Nandy25,Baiguera2025,Rabinovici25}, quantum thermodynamics, and quantum simulation \cite{Vasilyev20,Joshi22,Dong25} and information processing \cite{Zanardi21}. \\

{\it Acknowledgments.}--- It is a pleasure to thank Tadashi Takayanagi for insightful discussions and hospitality at the Yukawa Institute for Theoretical Physics at Kyoto University. This project was supported by the Luxembourg National Research Fund (FNR Grant Nos.\ C22/MS/17132054/AQCQNET and C24/MS/18940482/STAOpen). 

{\it Data Availability.}--- The data and source codes that support our findings are openly available \cite{data}.

\bibliography{SFFmonitoringv2}

@article{CaoXu22,
  title = {Probing quantum chaos in multipartite systems},
  author = {Cao, Zan and Xu, Zhenyu and del Campo, Adolfo},
  journal = {Phys. Rev. Res.},
  volume = {4},
  issue = {3},
  pages = {033093},
  numpages = {9},
  year = {2022},
  month = {Aug},
  publisher = {American Physical Society},
  doi = {10.1103/PhysRevResearch.4.033093},
  url = {https://link.aps.org/doi/10.1103/PhysRevResearch.4.033093}
}

@article{Lewalle18,
  title = {Chaos in continuously monitored quantum systems: An optimal-path approach},
  author = {Lewalle, Philippe and Steinmetz, John and Jordan, Andrew N.},
  journal = {Phys. Rev. A},
  volume = {98},
  issue = {1},
  pages = {012141},
  numpages = {21},
  year = {2018},
  month = {Jul},
  publisher = {American Physical Society},
  doi = {10.1103/PhysRevA.98.012141},
  url = {https://link.aps.org/doi/10.1103/PhysRevA.98.012141}
}

@article{Kruchkov23,
  title = {Engineering SYK Interactions in Disordered Graphene Flakes under Realistic Experimental Conditions},
  author = {Brzezi\ifmmode \acute{n}\else \'{n}\fi{}ska, Marta and Guan, Yifei and Yazyev, Oleg V. and Sachdev, Subir and Kruchkov, Alexander},
  journal = {Phys. Rev. Lett.},
  volume = {131},
  issue = {3},
  pages = {036503},
  numpages = {6},
  year = {2023},
  month = {Jul},
  publisher = {American Physical Society},
  doi = {10.1103/PhysRevLett.131.036503},
  url = {https://link.aps.org/doi/10.1103/PhysRevLett.131.036503}
}

@misc{Rabinovici25,
      title={Krylov Complexity}, 
      author={Eliezer Rabinovici and Adrián Sánchez-Garrido and Ruth Shir and Julian Sonner},
      year={2025},
      eprint={2507.06286},
      archivePrefix={arXiv},
      primaryClass={hep-th},
      url={https://arxiv.org/abs/2507.06286}, 
}

@misc{Baiguera2025,
      title={Quantum complexity in gravity, quantum field theory, and quantum information science}, 
      author={Stefano Baiguera and Vijay Balasubramanian and Pawel Caputa and Shira Chapman and Jonas Haferkamp and Michal P. Heller and Nicole Yunger Halpern},
      year={2025},
      eprint={2503.10753},
      archivePrefix={arXiv},
      primaryClass={hep-th},
      url={https://arxiv.org/abs/2503.10753}, 
}

@article{Nandy25,
title = {Quantum dynamics in Krylov space: Methods and applications},
journal = {Physics Reports},
volume = {1125-1128},
pages = {1-82},
year = {2025},
note = {Quantum dynamics in Krylov space: Methods and applications},
issn = {0370-1573},
doi = {https://doi.org/10.1016/j.physrep.2025.05.001},
url = {https://www.sciencedirect.com/science/article/pii/S0370157325001462},
author = {Pratik Nandy and Apollonas S. Matsoukas-Roubeas and Pablo Martínez-Azcona and Anatoly Dymarsky and Adolfo {del Campo}}
}

@article{Chowdhury22,
  title = {Sachdev-Ye-Kitaev models and beyond: Window into non-Fermi liquids},
  author = {Chowdhury, Debanjan and Georges, Antoine and Parcollet, Olivier and Sachdev, Subir},
  journal = {Rev. Mod. Phys.},
  volume = {94},
  issue = {3},
  pages = {035004},
  numpages = {78},
  year = {2022},
  month = {Sep},
  publisher = {American Physical Society},
  doi = {10.1103/RevModPhys.94.035004},
  url = {https://link.aps.org/doi/10.1103/RevModPhys.94.035004}
}

@article{GarciaALvarez17,
  title = {Digital Quantum Simulation of Minimal $\mathrm{AdS}/\mathrm{CFT}$},
  author = {Garc\'{\i}a-\'Alvarez, L. and Egusquiza, I. L. and Lamata, L. and del Campo, A. and Sonner, J. and Solano, E.},
  journal = {Phys. Rev. Lett.},
  volume = {119},
  issue = {4},
  pages = {040501},
  numpages = {6},
  year = {2017},
  month = {Jul},
  publisher = {American Physical Society},
  doi = {10.1103/PhysRevLett.119.040501},
  url = {https://link.aps.org/doi/10.1103/PhysRevLett.119.040501}
}

@book{Zurek25,
  author    = {Zurek, Wojciech Hubert},
  title     = {Decoherence and Quantum Darwinism: From Quantum Foundations to Classical Reality},
  publisher = {Cambridge University Press},
  address   = {Cambridge},
  year      = {2025}
}

@article{DAlessio16,
  author    = {Luca D'Alessio and Yariv Kafri and Anatoli Polkovnikov and Marcos Rigol},
  title     = {From quantum chaos and eigenstate thermalization to statistical mechanics and thermodynamics},
  journal   = {Advances in Physics},
  volume    = {65},
  number    = {3},
  pages     = {239--362},
  year      = {2016},
  publisher = {Taylor \& Francis},
  doi       = {10.1080/00018732.2016.1198134},
  url       = {https://doi.org/10.1080/00018732.2016.1198134}
}

@article{Asaduzzaman24,
  title = {Sachdev-Ye-Kitaev model on a noisy quantum computer},
  author = {Asaduzzaman, Muhammad and Jha, Raghav G. and Sambasivam, Bharath},
  journal = {Phys. Rev. D},
  volume = {109},
  issue = {10},
  pages = {105002},
  numpages = {7},
  year = {2024},
  month = {May},
  publisher = {American Physical Society},
  doi = {10.1103/PhysRevD.109.105002},
  url = {https://link.aps.org/doi/10.1103/PhysRevD.109.105002}
}

@Article{Luo2019,
author={Luo, Zhihuang
and You, Yi-Zhuang
and Li, Jun
and Jian, Chao-Ming
and Lu, Dawei
and Xu, Cenke
and Zeng, Bei
and Laflamme, Raymond},
title={Quantum simulation of the non-fermi-liquid state of Sachdev-Ye-Kitaev model},
journal={npj Quantum Information},
year={2019},
month={Jun},
day={18},
volume={5},
number={1},
pages={53},
issn={2056-6387},
doi={10.1038/s41534-019-0166-7},
url={https://doi.org/10.1038/s41534-019-0166-7}
}

@article{Danshita17,
    author = {Danshita, Ippei and Hanada, Masanori and Tezuka, Masaki},
    title = {Creating and probing the Sachdev–Ye–Kitaev model with ultracold gases: Towards experimental studies of quantum gravity},
    journal = {Progress of Theoretical and Experimental Physics},
    volume = {2017},
    number = {8},
    pages = {083I01},
    year = {2017},
    month = {08},
    issn = {2050-3911},
    doi = {10.1093/ptep/ptx108},
    url = {https://doi.org/10.1093/ptep/ptx108}}

@article{Matsoukas23SFFSA,
  title = {Unitarity breaking in self-averaging spectral form factors},
  author = {Matsoukas-Roubeas, Apollonas S. and Beau, Mathieu and Santos, Lea F. and del Campo, Adolfo},
  journal = {Phys. Rev. A},
  volume = {108},
  issue = {6},
  pages = {062201},
  numpages = {11},
  year = {2023},
  month = {Dec},
  publisher = {American Physical Society},
  doi = {10.1103/PhysRevA.108.062201},
  url = {https://link.aps.org/doi/10.1103/PhysRevA.108.062201}
}

@article{Vasilyev20,
  title = {Monitoring Quantum Simulators via Quantum Nondemolition Couplings to Atomic Clock Qubits},
  author = {Vasilyev, Denis V. and Grankin, Andrey and Baranov, Mikhail A. and Sieberer, Lukas M. and Zoller, Peter},
  journal = {PRX Quantum},
  volume = {1},
  issue = {2},
  pages = {020302},
  numpages = {22},
  year = {2020},
  month = {Oct},
  publisher = {American Physical Society},
  doi = {10.1103/PRXQuantum.1.020302},
  url = {https://link.aps.org/doi/10.1103/PRXQuantum.1.020302}
}

@article{Joshi22,
  title = {Probing Many-Body Quantum Chaos with Quantum Simulators},
  author = {Joshi, Lata Kh and Elben, Andreas and Vikram, Amit and Vermersch, Beno\^{\i}t and Galitski, Victor and Zoller, Peter},
  journal = {Phys. Rev. X},
  volume = {12},
  issue = {1},
  pages = {011018},
  numpages = {34},
  year = {2022},
  month = {Jan},
  publisher = {American Physical Society},
  doi = {10.1103/PhysRevX.12.011018},
  url = {https://link.aps.org/doi/10.1103/PhysRevX.12.011018}
}

@article{Dong25,
  title = {Measuring the Spectral Form Factor in Many-Body Chaotic and Localized Phases of Quantum Processors},
  author = {Dong, Hang and Zhang, Pengfei and Da\ifmmode \breve{g}\else \u{g}\fi{}, Ceren B. and Gao, Yu and Wang, Ning and Deng, Jinfeng and Zhang, Xu and Chen, Jiachen and Xu, Shibo and Wang, Ke and Wu, Yaozu and Zhang, Chuanyu and Jin, Feitong and Zhu, Xuhao and Zhang, Aosai and Zou, Yiren and Tan, Ziqi and Cui, Zhengyi and Zhu, Zitian and Shen, Fanhao and Li, Tingting and Zhong, Jiarun and Bao, Zehang and Li, Hekang and Wang, Zhen and Guo, Qiujiang and Song, Chao and Liu, Fangli and Chan, Amos and Ying, Lei and Wang, H.},
  journal = {Phys. Rev. Lett.},
  volume = {134},
  issue = {1},
  pages = {010402},
  numpages = {7},
  year = {2025},
  month = {Jan},
  publisher = {American Physical Society},
  doi = {10.1103/PhysRevLett.134.010402},
  url = {https://link.aps.org/doi/10.1103/PhysRevLett.134.010402}
}

@article{VallejoFabila24,
  title = {Reducing dynamical fluctuations and enforcing self-averaging by opening many-body quantum systems},
  author = {Vallejo-Fabila, Isa\'{\i}as and Das, Adway Kumar and Zarate-Herrada, David A. and Matsoukas-Roubeas, Apollonas S. and Torres-Herrera, E. Jonathan and Santos, Lea F.},
  journal = {Phys. Rev. B},
  volume = {110},
  issue = {7},
  pages = {075138},
  numpages = {12},
  year = {2024},
  month = {Aug},
  publisher = {American Physical Society},
  doi = {10.1103/PhysRevB.110.075138},
  url = {https://link.aps.org/doi/10.1103/PhysRevB.110.075138}
}

@article{Adway25,
  title = {Proposal for many-body quantum chaos detection},
  author = {Das, Adway Kumar and Cianci, Cameron and Cabral, Delmar G. A. and Zarate-Herrada, David A. and Pinney, Patrick and Pilatowsky-Cameo, Sa\'ul and Matsoukas-Roubeas, Apollonas S. and Batista, Victor S. and del Campo, Adolfo and Torres-Herrera, E. Jonathan and Santos, Lea F.},
  journal = {Phys. Rev. Res.},
  volume = {7},
  issue = {1},
  pages = {013181},
  numpages = {12},
  year = {2025},
  month = {Feb},
  publisher = {American Physical Society},
  doi = {10.1103/PhysRevResearch.7.013181},
  url = {https://link.aps.org/doi/10.1103/PhysRevResearch.7.013181}
}

@article{Alhassid1992,
  title = {Spectral autocorrelation function in the statistical theory of energy levels},
  author = {Alhassid, Y. and Levine, R. D.},
  journal = {Phys. Rev. A},
  volume = {46},
  issue = {8},
  pages = {4650--4653},
  numpages = {0},
  year = {1992},
  month = {Oct},
  publisher = {American Physical Society},
  doi = {10.1103/PhysRevA.46.4650},
  url = {http://link.aps.org/doi/10.1103/PhysRevA.46.4650}
}

@Article{Zhou2024,
author={Zhou, Yi-Neng
and Zhou, Tian-Gang
and Zhang, Pengfei},
title={General properties of the spectral form factor in open quantum systems},
journal={Frontiers of Physics},
year={2024},
month={Apr},
day={30},
volume={19},
number={3},
pages={31202},
issn={2095-0470},
doi={10.1007/s11467-024-1406-7},
url={https://doi.org/10.1007/s11467-024-1406-7}
}

@Article{MatsoukasRoubeas23,
author={Matsoukas-Roubeas, Apollonas S. and Roccati, Federico and Cornelius, Julien and Xu, Zhenyu and Chenu, Aur{\'e}lia and del Campo, Adolfo},
title={Non-Hermitian Hamiltonian deformations in quantum mechanics},
journal={JHEP},
year={2023},
month={Jan},
day={13},
volume={2023},
number={1},
pages={60},
issn={1029-8479},
doi={10.1007/JHEP01(2023)060},
url={https://doi.org/10.1007/JHEP01(2023)060}
}

@misc{SM,
  title = {See Supplemental Material for details},
  year = {2026}
}

@article{TorresHerrera20,
  title = {Self-averaging in many-body quantum systems out of equilibrium: Approach to the localized phase},
  author = {Torres-Herrera, E. Jonathan and De Tomasi, Giuseppe and Schiulaz, Mauro and P\'erez-Bernal, Francisco and Santos, Lea F.},
  journal = {Phys. Rev. B},
  volume = {102},
  issue = {9},
  pages = {094310},
  numpages = {13},
  year = {2020},
  month = {Sep},
  publisher = {American Physical Society},
  doi = {10.1103/PhysRevB.102.094310},
  url = {https://link.aps.org/doi/10.1103/PhysRevB.102.094310}
}

@article{Prange97,
  title = {The Spectral Form Factor Is Not Self-Averaging},
  author = {Prange, R. E.},
  journal = {Phys. Rev. Lett.},
  volume = {78},
  issue = {12},
  pages = {2280--2283},
  numpages = {0},
  year = {1997},
  month = {Mar},
  publisher = {American Physical Society},
  doi = {10.1103/PhysRevLett.78.2280},
  url = {https://link.aps.org/doi/10.1103/PhysRevLett.78.2280}
}

@article{Zanardi21,
  title = {Information scrambling and chaos in open quantum systems},
  author = {Zanardi, Paolo and Anand, Namit},
  journal = {Phys. Rev. A},
  volume = {103},
  issue = {6},
  pages = {062214},
  numpages = {16},
  year = {2021},
  month = {Jun},
  publisher = {American Physical Society},
  doi = {10.1103/PhysRevA.103.062214},
  url = {https://link.aps.org/doi/10.1103/PhysRevA.103.062214}
}

@article{Suntajs20,
  title = {Quantum chaos challenges many-body localization},
  author = {\ifmmode \check{S}\else \v{S}\fi{}untajs, Jan and Bon\ifmmode \check{c}\else \v{c}\fi{}a, Janez and Prosen, Toma\ifmmode \check{z}\else \v{z}\fi{} and Vidmar, Lev},
  journal = {Phys. Rev. E},
  volume = {102},
  issue = {6},
  pages = {062144},
  numpages = {12},
  year = {2020},
  month = {Dec},
  publisher = {American Physical Society},
  doi = {10.1103/PhysRevE.102.062144},
  url = {https://link.aps.org/doi/10.1103/PhysRevE.102.062144}
}

@Article{Gharibyan18,
author={Gharibyan, Hrant and Hanada, Masanori and Shenker, Stephen H. and Tezuka, Masaki},
title={Onset of random matrix behavior in scrambling systems},
journal={JHEP},
year={2018},
month={Jul},
day={18},
volume={2018},
number={7},
pages={124},
issn={1029-8479},
doi={10.1007/JHEP07(2018)124},
url={https://doi.org/10.1007/JHEP07(2018)124}
}

@Article{Orman2025,
author={Orman, Patrick
and Gharibyan, Hrant
and Preskill, John},
title={Quantum chaos in the sparse SYK model},
journal={Journal of High Energy Physics},
year={2025},
month={Feb},
day={26},
volume={2025},
number={2},
pages={173},
issn={1029-8479},
doi={10.1007/JHEP02(2025)173},
url={https://doi.org/10.1007/JHEP02(2025)173}
}

@ARTICLE{Chenu17work,
  author = {Aur\'elia Chenu and I{\~n}igo L. Egusquiza and Javier Molina-Vilaplana and Adolfo del Campo},
  title = {Quantum work statistics, Loschmidt echo and information scrambling},
  journal = {Sci. Rep.},
  year = {2018},
  volume = {8},
  pages = {12634},
  url = {https://doi.org/10.1038/s41598-018-30982-w}
}

@article{Chenu19,
  title = {Work statistics, loschmidt echo and information scrambling in chaotic quantum systems},
  author = {Chenu, Aur{\'e}lia and {Molina-Vilaplana}, Javier and {del Campo}, Adolfo},
  year = {2019},
  volume = {3},
  pages = {127},
  doi = {10.22331/q-2019-03-04-127},
  journal = {Quantum}
}

@Article{Cotler2017chaos,
author={Cotler, Jordan
and Hunter-Jones, Nicholas
and Liu, Junyu
and Yoshida, Beni},
title={Chaos, complexity, and random matrices},
journal={Journal of High Energy Physics},
year={2017},
month={Nov},
day={09},
volume={2017},
number={11},
pages={48},
issn={1029-8479},
doi={10.1007/JHEP11(2017)048},
url={https://doi.org/10.1007/JHEP11(2017)048}
}

@article{Can19,
	doi = {10.1088/1751-8121/ab4d26},
	url = {https://doi.org/10.1088%2F1751-8121%2Fab4d26},
	year = 2019,
	month = {nov},
	publisher = {{IOP} Publishing},
	volume = {52},
	number = {48},
	pages = {485302},
	author = {Tankut Can},
	title = {Random Lindblad dynamics},
	journal = {J. Phys. A: Math. Theor.},
}

@article{Can19b,
  title = {Spectral Gaps and Midgap States in Random Quantum Master Equations},
  author = {Can, Tankut and Oganesyan, Vadim and Orgad, Dror and Gopalakrishnan, Sarang},
  journal = {Phys. Rev. Lett.},
  volume = {123},
  issue = {23},
  pages = {234103},
  numpages = {6},
  year = {2019},
  month = {Dec},
  publisher = {American Physical Society},
  doi = {10.1103/PhysRevLett.123.234103},
  url = {https://link.aps.org/doi/10.1103/PhysRevLett.123.234103}
}

@article{Sa19,
  title = {Complex Spacing Ratios: A Signature of Dissipative Quantum Chaos},
  author = {S\'a, Lucas and Ribeiro, Pedro and Prosen, T.},
  journal = {Phys. Rev. X},
  volume = {10},
  issue = {2},
  pages = {021019},
  numpages = {23},
  year = {2020},
  month = {Apr},
  publisher = {American Physical Society},
  doi = {10.1103/PhysRevX.10.021019},
  url = {https://link.aps.org/doi/10.1103/PhysRevX.10.021019}
}

@article{SaProsen21,
  title = {Integrable nonunitary open quantum circuits},
  author = {S\'a, Lucas and Ribeiro, Pedro and Prosen, T.},
  journal = {Phys. Rev. B},
  volume = {103},
  issue = {11},
  pages = {115132},
  numpages = {9},
  year = {2021},
  month = {Mar},
  publisher = {American Physical Society},
  doi = {10.1103/PhysRevB.103.115132},
  url = {https://link.aps.org/doi/10.1103/PhysRevB.103.115132}
}

@article{SaProsen20,
  title = {Spectral transitions and universal steady states in random Kraus maps and circuits},
  author = {S{\'a}, L. and Ribeiro, P. and Can, T. and Prosen, T.},
  journal = {Phys. Rev. B},
  volume = {102},
  issue = {13},
  pages = {134310},
  numpages = {13},
  year = {2020},
  month = {Oct},
  publisher = {American Physical Society},
  doi = {10.1103/PhysRevB.102.134310},
  url = {https://link.aps.org/doi/10.1103/PhysRevB.102.134310}
}

@book{Carmichael09,
  title={Statistical Methods in Quantum Optics 2: Non-Classical Fields},
  author={Carmichael, H.J.},
  isbn={9783540713203},
  series={Theoretical and Mathematical Physics},
  year={2009},
  publisher={Springer Berlin Heidelberg}
}

@Article{delCampo2020,
author={del Campo, Adolfo and {Takayanagi}, Tadashi},
title={{Decoherence} in {Conformal} {Field} {Theory}},
journal={JHEP},
year={2020},
month={Feb},
day={26},
volume={2020},
number={2},
pages={170},
issn={1029-8479},
doi={10.1007/JHEP02(2020)170},
url={https://doi.org/10.1007/JHEP02(2020)170}
}

@book{BP02,
  title={The Theory of Open Quantum Systems},
  author={Breuer, H. P.  and Petruccione, F.},
  isbn={9780198520634},
  year={2002},
  publisher={Oxford University Press}
}

@article{Brody02,
    author = {Brody, Dorje C. and Hughston, Lane P.},
    title = {Efficient simulation of quantum state reduction},
    journal = {Journal of Mathematical Physics},
    volume = {43},
    number = {11},
    pages = {5254-5261},
    year = {2002},
    month = {11},
    issn = {0022-2488},
    doi = {10.1063/1.1512975},
    url = {https://doi.org/10.1063/1.1512975}
}

@Article{Cotler2017BH,
author="Cotler, Jordan S. and Gur-Ari, Guy and Hanada, Masanori and Polchinski, Joseph and Saad, Phil and Shenker, Stephen H. and Stanford, Douglas and Streicher, Alexandre and Tezuka, Masaki",
title="Black holes and random matrices",
journal="J. High Energy Phys.",
year="2017",
month="May",
day="22",
volume="2017",
number="5",
pages="118",
issn="1029-8479",
doi="10.1007/JHEP05(2017)118",
url="https://doi.org/10.1007/JHEP05(2017)118"
}

@article{Xu21SFF,
  title = {Thermofield dynamics: Quantum chaos versus decoherence},
  author = {Xu, Z. and Chenu, A. and Prosen, T. and del Campo, A.},
  journal = {Phys. Rev. B},
  volume = {103},
  issue = {6},
  pages = {064309},
  numpages = {11},
  year = {2021},
  month = {Feb},
  publisher = {American Physical Society},
  doi = {10.1103/PhysRevB.103.064309},
  url = {https://link.aps.org/doi/10.1103/PhysRevB.103.064309}
}

@article{Cornelius21,
  title = {Spectral Filtering Induced by Non-Hermitian Evolution with Balanced Gain and Loss: Enhancing Quantum Chaos},
  author = {Cornelius, Julien and Xu, Zhenyu and Saxena, Avadh and Chenu, Aur\'elia and del Campo, Adolfo},
  journal = {Phys. Rev. Lett.},
  volume = {128},
  issue = {19},
  pages = {190402},
  numpages = {6},
  year = {2022},
  month = {May},
  publisher = {American Physical Society},
  doi = {10.1103/PhysRevLett.128.190402},
  url = {https://link.aps.org/doi/10.1103/PhysRevLett.128.190402}
}

@article{Habib06,
  title = {Emergence of Chaos in Quantum Systems Far from the Classical Limit},
  author = {Habib, Salman and Jacobs, Kurt and Shizume, Kosuke},
  journal = {Phys. Rev. Lett.},
  volume = {96},
  issue = {1},
  pages = {010403},
  numpages = {4},
  year = {2006},
  month = {Jan},
  publisher = {American Physical Society},
  doi = {10.1103/PhysRevLett.96.010403},
  url = {https://link.aps.org/doi/10.1103/PhysRevLett.96.010403}
}

@article{Cucchietti03,
  title = {Decoherence and the Loschmidt Echo},
  author = {Cucchietti, F. M. and Dalvit, D. A. R. and Paz, J. P. and Zurek, W. H.},
  journal = {Phys. Rev. Lett.},
  volume = {91},
  issue = {21},
  pages = {210403},
  numpages = {4},
  year = {2003},
  month = {Nov},
  publisher = {American Physical Society},
  doi = {10.1103/PhysRevLett.91.210403},
  url = {https://link.aps.org/doi/10.1103/PhysRevLett.91.210403}
}

@article{Leviandier86,
  title = {Fourier Transform: A Tool to Measure Statistical Level Properties in Very Complex Spectra},
  author = {Leviandier, Luc and Lombardi, Maurice and Jost, R\'emi and Pique, Jean Paul},
  journal = {Phys. Rev. Lett.},
  volume = {56},
  issue = {23},
  pages = {2449--2452},
  numpages = {0},
  year = {1986},
  month = {Jun},
  publisher = {American Physical Society},
  doi = {10.1103/PhysRevLett.56.2449},
  url = {https://link.aps.org/doi/10.1103/PhysRevLett.56.2449}
}

@article{WilkieBrumer91,
  title = {Time-dependent manifestations of quantum chaos},
  author = {Wilkie, Joshua and Brumer, Paul},
  journal = {Phys. Rev. Lett.},
  volume = {67},
  issue = {10},
  pages = {1185--1188},
  numpages = {0},
  year = {1991},
  month = {Sep},
  publisher = {American Physical Society},
  doi = {10.1103/PhysRevLett.67.1185},
  url = {https://link.aps.org/doi/10.1103/PhysRevLett.67.1185}
}

@article{Ma95,
author = {Ma ,Jian-Zhong},
title = {Correlation Hole of Survival Probability and Level Statistics},
journal = {Journal of the Physical Society of Japan},
volume = {64},
number = {11},
pages = {4059-4063},
year = {1995},
doi = {10.1143/JPSJ.64.4059},
URL = {https://doi.org/10.1143/JPSJ.64.4059}
}

@book{Haake,
  title = {Quantum Signatures of Chaos},
  publisher = {Springer},
  year = {2010},
  author = {Haake, F.},
  address = {Berlin},
  url = {https://doi.org/10.1007/978-3-642-05428-0}
}

@article{Karkuszewski02,
  title = {Quantum Chaotic Environments, the Butterfly Effect, and Decoherence},
  author = {Karkuszewski, Zbyszek P. and Jarzynski, Christopher and Zurek, Wojciech H.},
  journal = {Phys. Rev. Lett.},
  volume = {89},
  issue = {17},
  pages = {170405},
  numpages = {4},
  year = {2002},
  month = {Oct},
  publisher = {American Physical Society},
  doi = {10.1103/PhysRevLett.89.170405},
  url = {https://link.aps.org/doi/10.1103/PhysRevLett.89.170405}
}

@article{Zurek03,
  title = {Decoherence, einselection, and the quantum origins of the classical},
  author = {Zurek, Wojciech Hubert},
  journal = {Rev. Mod. Phys.},
  volume = {75},
  issue = {3},
  pages = {715--775},
  numpages = {0},
  year = {2003},
  month = {May},
  publisher = {American Physical Society},
  doi = {10.1103/RevModPhys.75.715},
  url = {https://link.aps.org/doi/10.1103/RevModPhys.75.715}
}

@Article{MSS16,
author={Maldacena, Juan and Shenker, Stephen H. and Stanford, Douglas},
title={A bound on chaos},
journal={JHEP},
year={2016},
month={Aug},
day={17},
volume={2016},
number={8},
pages={106},
issn={1029-8479},
doi={10.1007/JHEP08(2016)106},
url={https://doi.org/10.1007/JHEP08(2016)106}
}

@Article{Cotler17,
author="Cotler, Jordan S. and Gur-Ari, Guy and Hanada, Masanori and Polchinski, Joseph and Saad, Phil and Shenker, Stephen H. and Stanford, Douglas and Streicher, Alexandre and Tezuka, Masaki",
title="Black holes and random matrices",
journal="JHEP",
year="2017",
month="May",
day="22",
volume="2017",
number="5",
pages="118",
issn="1029-8479",
doi="10.1007/JHEP05(2017)118",
url="https://doi.org/10.1007/JHEP05(2017)118"
}

@article{Egusquiza03,
  title = {Real clocks and the Zeno effect},
  author = {Egusquiza, I. L. and Garay, Luis J.},
  journal = {Phys. Rev. A},
  volume = {68},
  issue = {2},
  pages = {022104},
  numpages = {10},
  year = {2003},
  month = {Aug},
  publisher = {American Physical Society},
  doi = {10.1103/PhysRevA.68.022104},
  url = {https://link.aps.org/doi/10.1103/PhysRevA.68.022104}
}

@article{Egusquiza99,
  title = {Quantum evolution according to real clocks},
  author = {Egusquiza, I. L. and Garay, L. J. and Raya, J. M.},
  journal = {Phys. Rev. A},
  volume = {59},
  issue = {5},
  pages = {3236--3240},
  numpages = {0},
  year = {1999},
  month = {May},
  publisher = {American Physical Society},
  doi = {10.1103/PhysRevA.59.3236},
  url = {https://link.aps.org/doi/10.1103/PhysRevA.59.3236}
}

@article{Tsuji18,
  title = {Bound on the exponential growth rate of out-of-time-ordered correlators},
  author = {Tsuji, Naoto and Shitara, Tomohiro and Ueda, Masahito},
  journal = {Phys. Rev. E},
  volume = {98},
  issue = {1},
  pages = {012216},
  numpages = {7},
  year = {2018},
  month = {Jul},
  publisher = {American Physical Society},
  doi = {10.1103/PhysRevE.98.012216},
  url = {https://link.aps.org/doi/10.1103/PhysRevE.98.012216}
}

@article{Gisin84,
  title = {Quantum Measurements and Stochastic Processes},
  author = {Gisin, N.},
  journal = {Phys. Rev. Lett.},
  volume = {52},
  issue = {19},
  pages = {1657--1660},
  numpages = {0},
  year = {1984},
  month = {May},
  publisher = {American Physical Society},
  doi = {10.1103/PhysRevLett.52.1657},
  url = {https://link.aps.org/doi/10.1103/PhysRevLett.52.1657}
}

@article{Percival94,
author = {Percival, I. C.},
title = {Primary state diffusion},
journal = {Proceedings of the Royal Society of London. Series A: Mathematical and Physical Sciences},
volume = {447},
number = {1929},
pages = {189-209},
year = {1994},
doi = {10.1098/rspa.1994.0135},
URL = {https://royalsocietypublishing.org/doi/abs/10.1098/rspa.1994.0135}
}

@article{Diosi98,
  title = {Non-Markovian quantum state diffusion},
  author = {Di\'osi, L. and Gisin, N. and Strunz, W. T.},
  journal = {Phys. Rev. A},
  volume = {58},
  issue = {3},
  pages = {1699--1712},
  numpages = {0},
  year = {1998},
  month = {Sep},
  publisher = {American Physical Society},
  doi = {10.1103/PhysRevA.58.1699},
  url = {https://link.aps.org/doi/10.1103/PhysRevA.58.1699}
}

@article{Milburn91,
  title = {Intrinsic decoherence in quantum mechanics},
  author = {Milburn, G. J.},
  journal = {Phys. Rev. A},
  volume = {44},
  issue = {9},
  pages = {5401--5406},
  numpages = {0},
  year = {1991},
  month = {Nov},
  publisher = {American Physical Society},
  doi = {10.1103/PhysRevA.44.5401},
  url = {https://link.aps.org/doi/10.1103/PhysRevA.44.5401}
}

@article{Xu19,
  title = {Extreme Decoherence and Quantum Chaos},
  author = {Xu, Zhenyu and Garc\'{\i}a-Pintos, Luis Pedro and Chenu, Aur\'elia and del Campo, Adolfo},
  journal = {Phys. Rev. Lett.},
  volume = {122},
  issue = {1},
  pages = {014103},
  numpages = {6},
  year = {2019},
  month = {Jan},
  publisher = {American Physical Society},
  doi = {10.1103/PhysRevLett.122.014103},
  url = {https://link.aps.org/doi/10.1103/PhysRevLett.122.014103}
}

@article{MatsoukasRoubeas24PQC,
  doi = {10.22331/q-2024-08-27-1446},
  url = {https://doi.org/10.22331/q-2024-08-27-1446},
  title = {Quantum {C}haos and {C}oherence: {R}andom {P}arametric {Q}uantum {C}hannels},
  author = {Matsoukas-Roubeas, Apollonas S. and Prosen, Toma{\v{z}} and Campo, Adolfo del},
  journal = {{Quantum}},
  issn = {2521-327X},
  publisher = {{Verein zur F{\"{o}}rderung des Open Access Publizierens in den Quantenwissenschaften}},
  volume = {8},
  pages = {1446},
  month = aug,
  year = {2024}
}

@article{delcampo17,
  title = {Scrambling the spectral form factor: Unitarity constraints and exact results},
  author = {del Campo, A. and Molina-Vilaplana, J. and Sonner, J.},
  journal = {Phys. Rev. D},
  volume = {95},
  issue = {12},
  pages = {126008},
  numpages = {16},
  year = {2017},
  month = {Jun},
  publisher = {American Physical Society},
  doi = {10.1103/PhysRevD.95.126008},
  url = {https://link.aps.org/doi/10.1103/PhysRevD.95.126008}
}

@article{Adler03,
  title = {Weisskopf-Wigner decay theory for the energy-driven stochastic Schr\"odinger equation},
  author = {Adler, Stephen L.},
  journal = {Phys. Rev. D},
  volume = {67},
  issue = {2},
  pages = {025007},
  numpages = {14},
  year = {2003},
  month = {Jan},
  publisher = {American Physical Society},
  doi = {10.1103/PhysRevD.67.025007},
  url = {https://link.aps.org/doi/10.1103/PhysRevD.67.025007}
}

@book{Mehta_1991_Random_Matrices_Review,
  title={Random Matrices},
  author={Mehta, M.L.},
  isbn={9780124880511},
  lccn={90000257},
  url={https://books.google.lu/books?id=-sloQgAACAAJ},
  year={1991},
  publisher={Academic Press}
}

@book{Jacobs2014,
  title = {Quantum Measurement Theory and its Applications},
  ISBN = {9781139179027},
  url = {http://dx.doi.org/10.1017/CBO9781139179027},
  DOI = {10.1017/cbo9781139179027},
  publisher = {Cambridge University Press},
  author = {Jacobs,  Kurt},
  year = {2014},
  month = aug 
}

@book{Wiseman2009,
  title = {Quantum Measurement and Control},
  ISBN = {9781107424159},
  url = {http://dx.doi.org/10.1017/CBO9780511813948},
  DOI = {10.1017/cbo9780511813948},
  publisher = {Cambridge University Press},
  author = {Wiseman,  Howard M. and Milburn,  Gerard J.},
  year = {2009},
  month = nov 
}

@book{supp_ksendal2003,
  title = {Stochastic Differential Equations},
  ISBN = {9783642143946},
  ISSN = {2191-6675},
  url = {http://dx.doi.org/10.1007/978-3-642-14394-6},
  DOI = {10.1007/978-3-642-14394-6},
  journal = {Universitext},
  publisher = {Springer Berlin Heidelberg},
  author = {Øksendal,  Bernt},
  year = {2003}
}

@article{supp_Palmowski2002,
title = "A technique for exponential change of measure for Markov processes",
author = "Z.B. Palmowski and T. Rolski",
year = "2002",
volume = "8",
pages = "767--785",
journal = "Bernoulli",
issn = "1350-7265",
publisher = "Bernoulli Society for Mathematical Statistics and Probability",
number = "6",
url = {https://www.jstor.org/stable/3318901}
}

@book{supp_stnel2000,
  title = {Transformation of Measure on Wiener Space},
  ISBN = {9783662132258},
  ISSN = {1439-7382},
  url = {http://dx.doi.org/10.1007/978-3-662-13225-8},
  DOI = {10.1007/978-3-662-13225-8},
  journal = {Springer Monographs in Mathematics},
  publisher = {Springer Berlin Heidelberg},
  author = {\"{U}st\"{u}nel,  Ali S\"{u}leyman and Zakai,  Moshe},
  year = {2000}
}

@misc{data,
  doi = {10.5281/zenodo.18408585},
  url = {https://zenodo.org/doi/10.5281/zenodo.18408585},
  author = {Gopalakrishnan,  Preethi and Grabarits,  András József and del Campo,  Adolfo},
  title = {Data set and source codes for this article},
  publisher = {Zenodo},
  year = {2026},
  copyright = {Creative Commons Attribution 4.0 International}
}

@article{Okvatovity2020,
  author  = {Okv{\'a}tovity, Z. and D{\'o}ra, B.},
  title   = {Out-of-time-ordered commutators in {Dirac}-{Weyl} systems},
  journal = {Phys. Rev. B},
  volume  = {101},
  pages   = {245125},
  year    = {2020},
  doi     = {10.1103/PhysRevB.101.245125},
  eprint  = {1909.09376},
  archivePrefix = {arXiv},
}

@article{Hashimoto2017,
  author  = {Hashimoto, K. and Murata, K. and Yoshii, R.},
  title   = {Out-of-time-order correlators in quantum mechanics},
  journal = {J. High Energy Phys.},
  volume  = {2017},
  number  = {10},
  pages   = {138},
  year    = {2017},
  doi     = {10.1007/JHEP10(2017)138},
  eprint  = {1703.09435},
  archivePrefix = {arXiv},
}

@article{MaldacenaStanford2016,
  author  = {Maldacena, J. and Stanford, D.},
  title   = {Remarks on the {Sachdev}-{Ye}-{Kitaev} model},
  journal = {Phys. Rev. D},
  volume  = {94},
  pages   = {106002},
  year    = {2016},
  doi     = {10.1103/PhysRevD.94.106002},
  eprint  = {1604.07818},
  archivePrefix = {arXiv},
}

@misc{Kitaev2015,
  author       = {Kitaev, A.},
  title        = {A simple model of quantum holography},
  howpublished = {KITP Program: Entanglement in Strongly-Correlated Quantum Matter (Apr.~7,~2015 and May~27,~2015)},
  year         = {2015},
  note         = {\url{http://online.kitp.ucsb.edu/online/entangled15/kitaev/}, \url{http://online.kitp.ucsb.edu/online/entangled15/kitaev2/}},
}

@article{MaldacenaShenkerStanford2016,
  author  = {Maldacena, J. and Shenker, S. H. and Stanford, D.},
  title   = {A bound on chaos},
  journal = {J. High Energy Phys.},
  volume  = {2016},
  number  = {8},
  pages   = {106},
  year    = {2016},
  doi     = {10.1007/JHEP08(2016)106},
  eprint  = {1503.01409},
  archivePrefix = {arXiv},
}

@article{Sa2020,
  author  = {S\'a, L. and Ribeiro, P. and Prosen, T.},
  title   = {Complex Spacing Ratios: A Signature of Dissipative Quantum Chaos},
  journal = {Phys. Rev. X},
  volume  = {10},
  pages   = {021019},
  year    = {2020}
}

@article{GarciaGarcia2022,
  author  = {Garc\'ia-Garc\'ia, A. M. and S\'a, L. and Verbaarschot, J. J. M.},
  title   = {Symmetry Classification and Universality in Non-Hermitian Many-Body Quantum Chaos by the Sachdev-Ye-Kitaev Model},
  journal = {Phys. Rev. X},
  volume  = {12},
  pages   = {021040},
  year    = {2022}
}

@article{Li2021,
  author  = {Li, J. and Prosen, T. and Chan, A.},
  title   = {Spectral Statistics of Non-Hermitian Matrices and Dissipative Quantum Chaos},
  journal = {Phys. Rev. Lett.},
  volume  = {127},
  pages   = {170602},
  year    = {2021}
}

@article{GarciaMata2018,
  author  = {Garc\'ia-Mata, I. and Saraceno, M. and Jalabert, R. A. and Roncaglia, A. J. and Wisniacki, D. A.},
  title   = {Chaos Signatures in the Short and Long Time Behavior of the Out-of-Time Ordered Correlator},
  journal = {Phys. Rev. Lett.},
  volume  = {121},
  pages   = {210601},
  year    = {2018},
  doi     = {10.1103/PhysRevLett.121.210601}
}

@misc{Akemann2024,
  author = {Akemann, Gernot and Balducci, Federico and Chenu, Aur\'elia and P\"a{\ss}ler, Patricia and Roccati, Federico and Shir, Ruth},
  title  = {Two transitions in complex eigenvalue statistics: Hermiticity and integrability breaking},
  year   = {2024},
  eprint = {2409.10625},
  archivePrefix = {arXiv}
}

@misc{SenKumar2024,
  author = {Sen, Sunidhi and Kumar, Santosh},
  title  = {Exact and asymptotic dissipative spectral form factor for elliptic Ginibre unitary ensemble},
  year   = {2024},
  eprint = {2407.17148},
  archivePrefix = {arXiv}
}

@article{Haake1987,
  author  = {Haake, F. and Ku{\'s}, M. and Scharf, R.},
  title   = {Classical and quantum chaos for a kicked top},
  journal = {Z. Phys. B},
  volume  = {65},
  pages   = {381--395},
  year    = {1987}
}

@article{Ku1987,
  author  = {Ku{\'s}, M. and Scharf, R. and Haake, F.},
  title   = {Symmetry versus degree of level repulsion for kicked quantum systems},
  journal = {Z. Phys. B},
  volume  = {66},
  pages   = {129--134},
  year    = {1987}
}

@book{HaakeBook,
  author    = {Haake, F. and Gnutzmann, S. and Ku{\'s}, M.},
  title     = {Quantum Signatures of Chaos},
  edition   = {4th},
  publisher = {Springer},
  address   = {Cham},
  year      = {2018}
}

@article{Wang2004,
  author  = {Wang, X. and Ghose, S. and Sanders, B. C. and Hu, B.},
  title   = {Entanglement as a signature of quantum chaos},
  journal = {Phys. Rev. E},
  volume  = {70},
  pages   = {016217},
  year    = {2004}
}

@article{Chaudhury2009,
  author  = {Chaudhury, S. and Smith, A. and Anderson, B. E. and Ghose, S. and Jessen, P. S.},
  title   = {Quantum signatures of chaos in a kicked top},
  journal = {Nature},
  volume  = {461},
  pages   = {768--771},
  year    = {2009}
}

@article{Neill2016,
  author  = {Neill, C. and others},
  title   = {Ergodic dynamics and thermalization in an isolated quantum system},
  journal = {Nat. Phys.},
  volume  = {12},
  pages   = {1037--1041},
  year    = {2016}
}

@article{BGS1984,
  author  = {Bohigas, O. and Giannoni, M. J. and Schmit, C.},
  title   = {Characterization of chaotic quantum spectra and universality of level fluctuation laws},
  journal = {Phys. Rev. Lett.},
  volume  = {52},
  pages   = {1--4},
  year    = {1984}
}

@book{Mehta2004,
  author    = {Mehta, M. L.},
  title     = {Random Matrices},
  edition   = {3rd},
  publisher = {Elsevier},
  address   = {Amsterdam},
  year      = {2004}
}

@article{Dyson1962,
  author  = {Dyson, F. J.},
  title   = {The threefold way. {A}lgebraic structure of symmetry groups and ensembles in quantum mechanics},
  journal = {J. Math. Phys.},
  volume  = {3},
  pages   = {1199--1215},
  year    = {1962}
}

@article{Brezin1997,
  author  = {Br{\'e}zin, E. and Hikami, S.},
  title   = {Spectral form factor in a random matrix theory},
  journal = {Phys. Rev. E},
  volume  = {55},
  pages   = {4067--4083},
  year    = {1997}
}

@article{Cotler2017,
  author  = {Cotler, J. S. and Gur-Ari, G. and Hanada, M. and Polchinski, J. and Saad, P. and Shenker, S. H. and Stanford, D. and Streicher, A. and Tezuka, M.},
  title   = {Black holes and random matrices},
  journal = {J. High Energy Phys.},
  volume  = {2017},
  number  = {5},
  pages   = {118},
  year    = {2017}
}

@article{delCampo2017,
  author  = {del Campo, A. and Molina-Vilaplana, J. and Sonner, J.},
  title   = {Scrambling the spectral form factor: {U}nitarity constraints and exact results},
  journal = {Phys. Rev. D},
  volume  = {95},
  pages   = {126008},
  year    = {2017}
}

@article{Xu2021,
  author  = {Xu, Z. and Chenu, A. and Prosen, T. and del Campo, A.},
  title   = {Thermofield dynamics: {Q}uantum chaos versus decoherence},
  journal = {Phys. Rev. B},
  volume  = {103},
  pages   = {064309},
  year    = {2021}
}

@article{Cornelius2022,
  author  = {Cornelius, J. and Xu, Z. and Saxena, A. and Chenu, A. and del Campo, A.},
  title   = {Spectral filtering induced by non-{H}ermitian evolution with balanced gain and loss: {E}nhancing quantum chaos},
  journal = {Phys. Rev. Lett.},
  volume  = {128},
  pages   = {190402},
  year    = {2022}
}

\twocolumngrid

\onecolumngrid
\clearpage
\vspace{1em}
\begin{center}
    \textbf{End Matter}
\end{center}
\twocolumngrid

\emph{Out-of-time-order correlator under continuous monitoring.}---
Here, we ask how continuous monitoring modifies another dynamical signature of quantum chaos, the operator growth. To this end, we study the out-of-time-order correlator (OTOC) of the monitored SYK model. 
For the coherent Gibbs state
$\ket{\psi_\beta},$ and
two distinct Majorana operators $\chi_i,\,\chi_j,\,i\neq j$, we define the normalized OTOC, 
\begin{equation}\label{Eq:OTOCdef}
\mathcal F_\beta(t)=\frac{F_\beta(t)}{F_\beta(0)},\quad
F_\beta(t)=\frac{\bra{\psi_\beta}\widetilde\chi_i(t)\,\chi_j\,
\widetilde\chi_i(t)\,\chi_j\ket{\psi_\beta}}{\mathcal N(t,W_t)},
\end{equation}
with the monitored Heisenberg evolution $\widetilde\chi_{i,nm}(t)=(\chi_i)_{nm}\,e^{i(E_n-E_m)t-2\gamma t(E_n^2+E_m^2)+\sqrt{2\gamma}W_t(E_n+E_m)}$
and the trajectory normalization $\mathcal N(t,W_t)$. By
construction $\mathcal F_\beta(0)=1$, and its decay from unity measures the growth of operator complexity. 

In the absence of monitoring, under unitary evolution, the OTOC initially departs from unity, quadratically~\cite{Hashimoto2017,GarciaMata2018,Okvatovity2020} and subsequently enters the exponential scrambling regime, $1-\mathcal F_\beta(t)\simeq\tfrac{c}{N}e^{\lambda_L t}$ for
$t\gtrsim1/J$~\cite{MaldacenaStanford2016,Kitaev2015,MaldacenaShenkerStanford2016}. Continuous monitoring introduces an additional decay channel, as the energy filter suppresses the correlator at a rate $\Gamma\simeq4\gamma\,\mathrm{Var}_\beta(E)$. The onset of the scrambling regime is determined by the interplay between the monitoring-induced contribution and the underlying operator growth,
\begin{equation}\label{Eq:OTOCdecay}
1-\mathcal F_\beta(t)\;\approx\;\Gamma\,t
+\frac{c}{N}\,e^{\lambda_L t}.
\end{equation}
Thus, increasing the monitoring strength does not simply modify the scrambling dynamics but also produces an increasingly pronounced departure from the unmonitored ($\gamma=0$) early time behavior, shifting the onset of the chaotic growth, as illustrated in Fig.~\ref{Fig:OTOC_tstar}(a).
\begin{figure}[t]
    \centering
    \includegraphics[width=1\columnwidth]{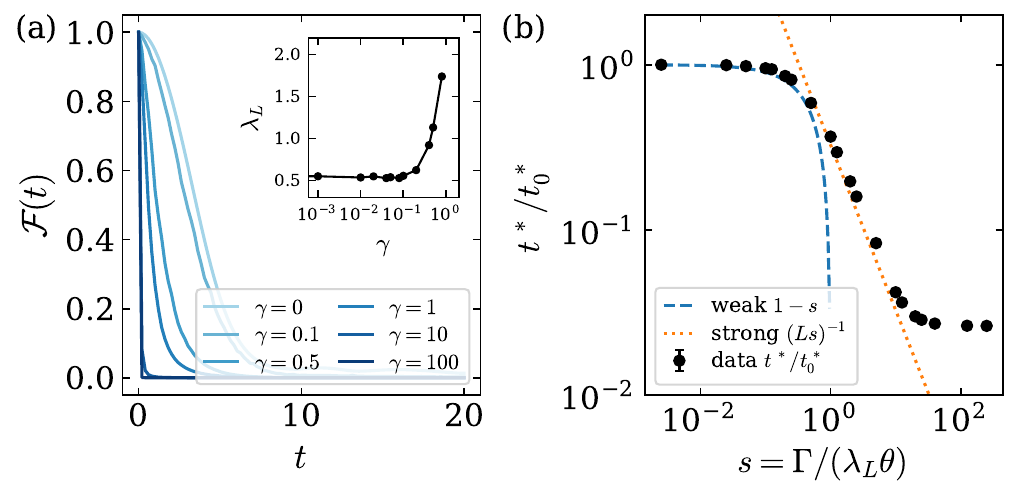}
    \caption{Tuning operator scrambling through continuous monitoring.
    (a) Evolution of the normalized OTOC
    $\mathcal F_\beta(t)=F_\beta(t)/F_\beta(0)$ for the monitored
    SYK model $(N=22, \beta=0)$, averaged over $100$ stochastic
    trajectories and SYK disorder. Increasing $\gamma$ progressively
    shortens the scrambling window through measurement-induced decay
    at the rate $\Gamma\simeq4\gamma\,\mathrm{Var}_\beta(E)$.
    The inset shows the fitted growth exponent
    $\lambda_{\rm eff}(\gamma)$, which remains near its intrinsic
    value $\simeq0.56\,J$ for $\gamma\lesssim0.05$ and is
    subsequently enhanced by the filter contribution of
    Eq.~\eqref{Eq:OTOCdecay}. (b) Rescaled deviation time
    $t^*/t^*_0$ versus $s=\Gamma/(\lambda_L\theta)$, showing the
    crossover from the weak-measurement behavior $1-s$ to the
    strong-measurement regime $(Ls)^{-1}$, followed by saturation
    at the time-grid resolution, with $L=\lambda_L t^*_0$.}
\label{Fig:OTOC_tstar}
\end{figure}
This provides a dynamical counterpart to the tailoring observed in the SFF. In the SFF, the energy filter suppresses the non-universal contributions of the spectral density and advances the onset of the ramp. In the OTOC, the same filter advances the departure from the unmonitored early time behavior and brings the scrambling regime to earlier times. The measurement strength therefore acts as a control parameter for the dynamical manifestation of Hamiltonian quantum chaos. 

For weak monitoring, the exponential scrambling window remains well defined. The additional filter contribution steepens the apparent growth of $1-\mathcal F_\beta$, so that the exponent $\lambda_{\rm eff}$ extracted from a simple exponential fit exceeds $\lambda_L$ and grows with $\gamma$ even though the underlying Lyapunov exponent remains unchanged at this order. The more direct signature of the tailoring is instead the shift of the crossover time $t^*$, which marks the onset of the exponential regime. For moderate $\gamma$, this time decreases approximately linearly with the filter strength, 
\begin{equation}\label{Eq:tstarweak}
t^*(\gamma)\;\approx\;t^*_0\big(1-s\big),\qquad
s=\frac{\Gamma}{\lambda_L\,\theta},
\end{equation}
with $t^*_0$ the unmonitored value and $\theta$ the
fractional margin by which $1-\mathcal F_\beta$ must depart from its fitted log-linear growth (see~\cite{SM} for details) . The crossover becomes particularly pronounced when $\Gamma$ becomes comparable to $\lambda_L$. At this point, the monitoring-induced decay competes directly with the intrinsic operator growth, and the scrambling window progressively disappears. For stronger monitoring, the OTOC decay is dominated by the energy filter rather than by operator growth, giving
\begin{equation}\label{Eq:tstar}
\frac{t^*}{t^*_0}\;\simeq\;\frac{1}{L\,s},\qquad
L=\lambda_L t^*_0 ,
\end{equation}
until the decay becomes faster than the resolution of the time grid, beyond which $t^*$ saturates at a finite value, as seen in Fig~\ref{Fig:OTOC_tstar}(b).

As in the SFF, continuous monitoring reshapes the timescales on which quantum chaos manifest, the strongest modification occurring when the monitoring-induced timescale becomes comparable to the intrinsic scrambling timescale, $\gamma\sim\mathcal O(1)$.\\

\emph{Dissipative Liouvillian spectrum.---} A complementary characterization of chaos in an open quantum system is provided by the spectral statistics of its Liouvillian generator. The same monitoring protocol that tailors the dynamical signatures of chaos in the SFF and OTOC maps the underlying Hamiltonian spectrum into the complex plane, providing a statistical view of how chaotic correlations are altered by the open-system dynamics.

\begin{figure}
\centering
\includegraphics[width=1\columnwidth]{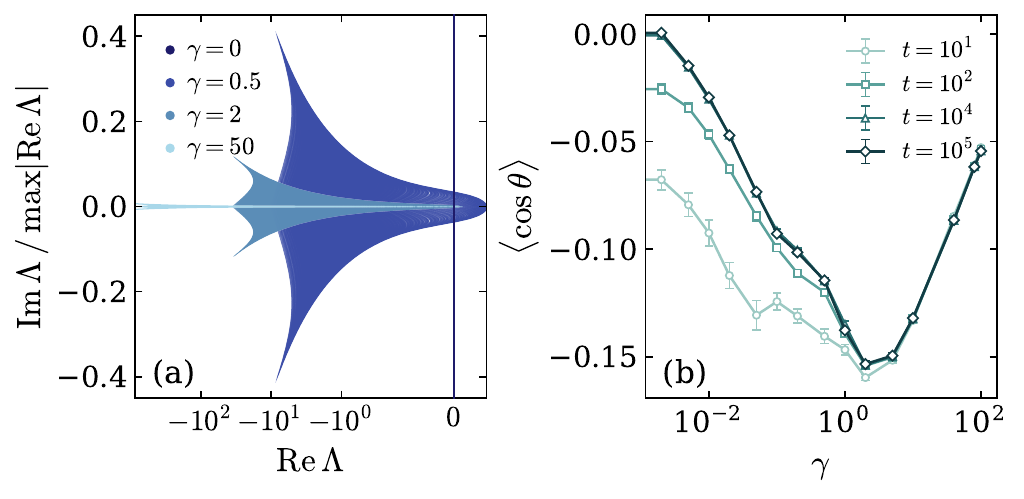}
    \caption{Dissipative Liouvillian spectrum of the monitored SYK model ($N=22$, $\beta=0$). (a) Trajectory-resolved scatter of $\Lambda_{nm}$ for $\gamma=0,0.5,2,50$, on a signed logarithmic real axis with the imaginary part rescaled by the real span $\max|\Re\Lambda|$ of each cloud. The quasi-one-dimensional spectrum at $\gamma=0$ (vertical line) deforms into a two-dimensional cloud at intermediate $\gamma$ and is stretched back along the real axis at large $\gamma$ (b) Angular order parameter $\langle\cos\theta\rangle$ versus $\gamma$ for several stroboscopic times $t$, averaged over disorder and noise, with the Poisson and GinUE references dashed. The minimum at $\gamma_{\rm opt}\sim\mathcal O(1)$ reaches the value attained by a GUE spectrum under the same map, and its position and depth are independent of $t$.}
\label{Fig:Liouville}
\end{figure}

When monitoring the energy of the system, the Liouvillian remains diagonal in the energy basis and each matrix element of the density matrix $\rho_{nm}(t)$ evolves independently for a given measurement record. Writing $\rho_{nm}(t)=e^{\Lambda_{nm}t}\rho_{nm}(0)$, its finite-time trajectory eigenvalue is
\begin{equation}\label{Eq}
\begin{aligned}
\Lambda_{nm}
=&\frac{1}{t}\log\frac{\rho_{nm}(t)}{\rho_{nm}(0)}\\
=&-i(E_n-E_m)
-\gamma\left[(E_n-E_m)^2+(E_n+E_m)^2\right] \\
&+\sqrt{2\gamma}\,(E_n+E_m)\frac{W_t}{t}.
\end{aligned}
\end{equation}
For timescales of the ramp, however, the additional stochastic term is negligible, decaying as $W_t/t\sim t^{-1/2}$, again showing that the Liouvillian spectrum for a typical trajectory exhibits similar features compared to the no-jump limit.

For each realization of $W_t$, the real and imaginary parts of the $d^2$ complex eigenvalues trace a two-dimensional locus in the complex plane. The real part is determined by $\gamma\big[(E_m-E_n)^2+(E_m+E_n)^2\big]$, while the imaginary part by $(E_m-E_n)$. Figure~\ref{Fig:Liouville}(a) shows the deformation of the Liouvillian eigenvalues on the complex plane as $\gamma$ is varied. At $\gamma=0$ the spectrum is quasi-one-dimensional, confined to the imaginary axis.
Monitoring deforms it into a genuinely two-dimensional ``cloud'' at
intermediate $\gamma$. For large $\gamma$, $\gamma\big[(E_m-E_n)^2+(E_m+E_n)^2\big]$ then deforms it along the real axis, returning it to a quasi-one-dimensional shape.

We characterize the spectrum by the complex spacing ratios~\cite{Sa2020},
\begin{equation}\label{Eq:CSR}
z_\alpha=\frac{\Lambda_\alpha^{\rm NN}-\Lambda_\alpha}
{\Lambda_\alpha^{\rm NNN}-\Lambda_\alpha}
=|z_\alpha|e^{i\theta_\alpha},
\qquad |z_\alpha|\le 1,
\end{equation}
with $\Lambda_\alpha^{\rm NN}$ and $\Lambda_\alpha^{\rm NNN}$ the nearest
and next-nearest neighbors in Euclidean distance, and the scalar order parameters
$\langle|z|\rangle$ and $\langle\cos\theta\rangle$ averaged over the
spectrum, the Wiener and the SYK realizations. For non-chaotic spectra, the statistical properties of the spectrum become identical to those of a two-dimensional Poisson statistics, characterized by $\langle\cos\theta\rangle=0$, while for chaotic spectral correlations the Ginibre unitary ensemble (GinUE) provides the reference value,
$\langle\cos\theta\rangle\simeq-0.241$.

In terms of the mean level spacings, complex spacing ratios are primarily governed by the difference $E_m-E_n$, which enters both the real and imaginary parts of the Liouvillian spectrum. The small-gap condition selects neighboring levels, while the corresponding sums $E_m+E_n$ connect levels that are typically far apart in the spectrum and effectively uncorrelated. The resulting nearest-neighbor distances consequently retain a Poisson-like character, pinning $\langle|z|\rangle$ near $2/3$ for all $\gamma$.

The relative orientation of neighboring eigenvalues, however, remains sensitive to the correlations inherited from the underlying Hamiltonian spectrum. Thus, while the radial statistics lose the signatures of level repulsion, the angular correlations retain them, making $\langle\cos\theta\rangle$ an efficient measure of chaotic structure in the Liouvillian spectrum. As shown in Fig.~\ref{Fig:Diptime}(b), it develops a minimum $\langle\cos\theta\rangle\simeq-0.16$ at $\gamma_{\rm opt}\sim\mathcal O(1)$, approaching the value obtained from a GUE spectrum under the same Liouvillian construction~\cite{Sa2020,GarciaGarcia2022}. The angular correlations therefore exhibit the same nonmonotonic dependence on $\gamma$ as the correlation hole in the SFF, reaching the two-dimensional Poissonian limit as $\gamma\to0$ and $\gamma\to\infty$, where the spectrum becomes effectively one-dimensional. Here, the phases alternate randomly between $\cos\theta=\pm1$, averaging to zero~\cite{Akemann2024}. The curves obtained at
different times converge with increasing $t$, as expected
from the $t^{-1/2}$ suppression of the noise term. 

To probe correlations beyond nearest neighbors, we further consider the dissipative SFF (DSFF), which resolves the complex spectrum along different directions in the complex plane. The DSFF shows the same enhancement of the correlation ramp with increasing $\gamma$ as the ordinary SFF, with the longest ramp obtained at intermediate monitoring strengths and progressively weaker dependence on $\gamma$ as the projection approaches the imaginary axis (see~\cite{SM}).
\appendix

\newpage
\clearpage

\title{---Supplementary Material---Quantum Chaos Under Continuous Quantum Measurements 
}
\maketitle
\onecolumngrid
\begin{center}
\textbf{\large Supplemental Material for\\
	    ``Tailoring Quantum Chaos With Continuous Quantum Measurements''}\\
\vspace{0.5cm}
Preethi Gopalakrishnan, Andr\'as Grabarits and Adolfo del Campo 
\end{center}
\renewcommand{\theequation}{S\arabic{equation}}
\renewcommand{\thefigure}{S\arabic{figure}}
\renewcommand{\thetable}{S\arabic{table}}

\setcounter{equation}{0}
\setcounter{figure}{0}
\setcounter{table}{0}
\setcounter{page}{1}

\onecolumngrid

\renewcommand{\theequation}{S\arabic{equation}}
\renewcommand{\thefigure}{S\arabic{figure}}
\renewcommand{\thetable}{S\arabic{table}}
\renewcommand{\bibnumfmt}[1]{[S#1]}

\setcounter{equation}{0}
\setcounter{figure}{0}
\setcounter{table}{0}
\setcounter{page}{1}

\section{Recalling Itô Calculus}
In this section, we summarize the basic rules of Itô calculus used throughout this work. 
A Wiener process $W_t=\int_0^t\mathrm dW_s$ is composed of independent increments following Gaussian distributions with $\mathbb{E}[\mathrm d W_t] = 0$ and $\mathbb E[(\mathrm{d}W_t)^2] = \mathrm d t.$ As a consequence, the Itô multiplication rules are 
\begin{equation}
    (\mathrm d t)^2 = 0, \qquad \mathrm d t \mathrm d W_t = 0, \qquad (\mathrm dW_t)^2 = \mathrm dt.
\end{equation}
For any stochastic process $X_t$ governed by a stochastic differential equation, 
\begin{equation}
    \mathrm dX_t = a_t\mathrm{d}t + b_t\mathrm{d}W_t,\quad (\mathrm dX_t)^2=b^2_t \mathrm dt,
\end{equation}
the stochastic derivative of a function $f(X_t,t)$, assumed to be twice continuously differentiable, reads
\begin{equation}
    \mathrm df(X_t,t)=\frac{\partial f}{\partial t}\mathrm dt+\frac{\partial f}{\partial X_t}\mathrm dX_t+\frac{1}{2}\frac{\partial^2f}{\partial X^2_t}(\mathrm dX_t)^2=\left(\frac{\partial f}{\partial t}+ a\,\frac{\partial f}{\partial X_t}+ \frac{1}{2} b^2\,\frac{\partial^2 f}{\partial X^2_t}\right) \mathrm dt+ b\,\frac{\partial f}{\partial X_t}\, \mathrm dW_t.
\end{equation}
This is known as the Itô's formula. 

As a direct consequence of Itô's formula, exponentiation of a stochastic process acquires an additional drift term. For $\mathrm d X_t = a X_t\mathrm{d} t + g X_t \mathrm{d}W_t$,
\begin{equation}\label{SuppEq:Exp}
    X_t = X_0\exp\left[\left(a-\frac{1}{2}g^2\right)t + gW_t \right].
\end{equation}
The product of two Itô processes $X_t$ and $Y_t$ satisfies the Itô product rule 
\begin{equation}
    \mathrm{d}(X_tY_t) = X_t \mathrm{d}Y_t + Y_t\mathrm{d}X_t + \mathrm{d}X_t \mathrm{d}Y_t.
\end{equation}
Similarly,  the ratio $R_t = X_t / Y_t$ satisfies

\begin{equation} \label{Eq:ItoQuotient}
\mathrm d\!\left(\frac{X_t}{Y_t}\right)
= \frac{1}{Y_t}\,\mathrm d X_t
- \frac{X_t}{Y_t^2}\,\mathrm d Y_t
+ \frac{X_t}{Y_t^3}(\mathrm d Y_t)^2
- \frac{1}{Y_t^2}\,\mathrm d X_t\,\mathrm d Y_t .
\end{equation}

\section{Equivalence of the dynamics under stochastic energy-diffusion and continuous energy monitoring}
The energy-driven stochastic Schr\"odinger equation has been extensively investigated in the foundations of physics \cite{Gisin84,Milburn91,Percival94,Diosi98,Adler03}, exploring quantum measurement theory and possible deviations of quantum mechanics, e.g., in the context of collapse models for state reduction. In this section, we establish its mathematical equivalence with the stochastic master equation (SME) describing continuous quantum measurements of the energy. 
For a pure state $|\psi\rangle$, the energy-driven stochastic Schr\"odinger equation takes the form
\begin{equation}
{\rm d}|\psi\rangle=-iH|\psi\rangle {\rm d} t-\gamma(H-\langle H\rangle)^2|\psi\rangle {\rm d} t+\sqrt{2\gamma}  (H-\langle H\rangle)|\psi\rangle {\rm d}W_t,
\end{equation}
where $\gamma$ controls the strength of the stochasticity and $\mathrm dW_t$ denotes the infinitesimal increment of the Wiener noise, $\mathbb E[\mathrm dW_t]=0$, $\mathbb E[\mathrm dW_t\mathrm dW_s]=\delta_{t,s}\mathrm dt$.
Note that this equation is nonlinear in the quantum state as the mean energy is given by $\langle H\rangle=\langle\psi|H|\psi\rangle$. 

Using It${\rm \hat{o}}$ calculus, to $\mathcal{O}({\rm d}t)$, the equation of motion for the pure density matrix $\rho=|\psi\rangle\langle\psi|$ is given by ${\rm d}\rho={\rm d}|\psi\rangle\langle\psi|+|\psi\rangle{\rm d}\langle\psi|+{\rm d}|\psi\rangle{\rm d}\langle\psi|$, which yields by direct computation
\begin{eqnarray}
{\rm d}\rho&=&-i[H,\rho]{\rm d}t-\gamma\left(\{(H-\langle H\rangle)^2,\rho\}-2(H-\langle H\rangle)\rho(H-\langle H\rangle) \right){\rm d}t+\sqrt{2\gamma}\{H-\langle H\rangle,\rho\}{\rm d}W_t\\
&=&-i[H,\rho]{\rm d}t-\gamma[H-\langle H\rangle,[H-\langle H\rangle,\rho]]{\rm d}t+\sqrt{2\gamma}\{H-\langle H\rangle,\rho\}{\rm d}W_t.
\end{eqnarray}
Simplifying the double commutator and rewriting the stochastic term, one obtains
\begin{equation}\label{QEMeq}
{\rm d}\rho=-i[H,\rho]{\rm d}t-\gamma[H,[H,\rho]{\rm d}t+\sqrt{2\gamma}\left(\{H,\rho\}-2\langle H\rangle\rho\right){\rm d}W_t, 
\end{equation}
which takes the form of the SME governing the evolution under continuous energy monitoring with unit detection efficiency, discussed in the main text. 
Equation (\ref{QEMeq}) corrects the expression previously quoted in the literature [e.g., Eq. (4c) in \cite{Adler03}].

\section{Solution to the Stochastic Master Equation}
We consider the SME describing continuous measurement of an observable $H$ with measurement strength $\gamma$,
\begin{equation}
\mathrm{d}\rho(t) = -i[H,\rho(t)]\,\mathrm{d}t - \gamma[H,[H,\rho(t)]]\,\mathrm{d}t + \sqrt{2\gamma}\,\big(\{H,\rho(t)\}-2\langle H\rangle\rho(t)\big)\,\mathrm{d}W_t,
\end{equation}
where $\langle H\rangle = \mathrm{Tr}[H \rho(t)]$. To linearize this nonlinear SME, we introduce an unnormalized density operator $\hat\rho(t)$ evolving according to the linear SME
\begin{equation}
\mathrm{d}\hat\rho(t) = -i[H,\hat\rho(t)]\,\mathrm{d}t - \gamma[H,[H,\hat\rho(t)]]\,\mathrm{d}t + \sqrt{2\gamma}\,\{H,\hat\rho(t)\}\,\mathrm{d}W_t,\label{linSME}
\end{equation}

Solving Eq.~(\ref{linSME}) in the energy eigenbasis, one finds
\begin{equation}\label{SuppEq:Rho_t}
    \rho(t) = \frac{ \sum_{nm}\rho_{nm}(0)e^{-i(E_n -E_m)t - 2\gamma t(E_n^2 + E_m^2) + \sqrt{2\gamma}W_t(E_n+E_m)}}{ \sum_k \rho_{kk}(0)e^{-4\gamma tE_k^2 + 2\sqrt{2\gamma}W_tE_k}}\lvert n\rangle \langle m\rvert.
\end{equation}

Here, the Itô correction has been properly accounted for using the identity in Eq.~\ref{SuppEq:Exp}. The nonlinear SME is thus mapped onto a linear stochastic evolution supplemented by a stochastic normalization factor.
Applying Itô's quotient rule Eq.~\eqref{Eq:ItoQuotient}, the evolution of the normalized state $\rho(t) = \hat{\rho}(t)/\mathrm{Tr}(\hat{\rho}(t)$, yields
\begin{equation}
\mathrm{d}\rho(t) = \frac{\mathrm{d}\hat\rho(t)}{\mathrm{Tr}[\hat\rho(t)]} - \frac{\hat\rho(t)\, \mathrm{Tr}[\mathrm{d}\hat\rho(t)]}{\mathrm{Tr}[\hat\rho(t)]^2} + \frac{\hat\rho(t)\, (\mathrm{Tr}[\mathrm{d}\hat\rho(t)])^2}{\mathrm{Tr}[\hat\rho(t)]^3} - \frac{\mathrm{d}\hat\rho(t)\, \mathrm{Tr}[\mathrm{d}\hat\rho(t)]}{\mathrm{Tr}[\hat\rho(t)]^2}.
\end{equation}
Substituting the linear SME and simplifying gives
\begin{equation}
\mathrm{d}\rho(t) = -\gamma [H,[H,\rho]]\,\mathrm{d}t + \sqrt{2\gamma}(\{H,\rho\}-2\langle H\rangle \rho)\,\mathrm{d}W_t - 4\gamma (\{H,\rho\}-2\langle H\rangle\rho) \langle H\rangle\,\mathrm{d}t,
\end{equation}
where the last term arises from stochastic normalization and Itô corrections. This deterministic drift can be absorbed into a shifted Wiener increment, using the Girsanov transformation \cite{supp_ksendal2003,supp_stnel2000,supp_Palmowski2002}. Specifically, defining $\mathrm{d}\widetilde W_t = \mathrm{d}W_t - 2\sqrt{2\gamma}\langle H\rangle\,\mathrm{d}t$, the drift term is canceled. The process $\lambda_t = 2 \sqrt{2\gamma}\langle H\rangle_t$ is fully determined by the noise history up to time $t$. Consequently, it can be considered a modified Wiener process $\widetilde{W}_t$.

Substituting $\mathrm{d}W_t = \mathrm{d}\widetilde W_t + 2\sqrt{2\gamma}\langle H\rangle \mathrm{d}t$ cancels the extra deterministic term, leaving the normalized nonlinear SME 
\begin{equation}
\mathrm{d}\rho(t) = -i[H,\rho]\,\mathrm{d}t - \gamma[H,[H,\rho]]\,\mathrm{d}t + \sqrt{2\gamma} (\{H,\rho\}-2\langle H\rangle \rho)\, \mathrm{d}\widetilde W_t.
\end{equation}

While the Wiener increment $\mathrm{d}W_t$ is a Gaussian random variable with probability density function $\exp[-(\mathrm{d}W_t)^2/(2\mathrm{d}t)]$, in the distribution of the shifted increment, $\mathrm{d}\widetilde W_t = \mathrm{d}W_t - \lambda_t\mathrm{d}t$ with $\lambda_t=\sqrt{8\gamma}\langle H\rangle_t$, the Gaussian weight can be rewritten as 
\begin{eqnarray}
    e^{-\mathrm dW^2_t/(2\mathrm dt)}=e^{-(\mathrm d W_t+\sqrt{8\gamma}\langle H\rangle dt)^2/(2\mathrm dt)}e^{\lambda^2_t dt/2}e^{\lambda_t dW_t}\equiv e^{-d\widetilde W^2_t/(2\mathrm dt)}e^{\lambda^2_t dt/2}e^{\lambda_t \mathrm dW_t}.
\end{eqnarray}

This means each Gaussian increment can be reparametrized in terms of $\widetilde{\mathrm{d}W_t}$ with the additional Jacobian, $e^{\lambda^2_t dt/2}e^{\lambda_t dW_t}$ in agreement with the Girsanov theorem. As a result, the statistical features of any arbitrary function $f(W_t-\int_0^t\mathrm ds\lambda_s)$ can always be described by the modified function $f(\widetilde W_t)e^{\frac{1}{2}\int_0^t\mathrm ds\lambda^2_s + \int_0^t\mathrm dW_s\lambda_s}$, where $\widetilde W_t$ acts as a standard Wiener noise. The products of $e^{-\mathrm d\widetilde W^2_t/2\mathrm dt}e^{\lambda^2_t dt/2}e^{\lambda_t \mathrm dW_t}$ for each $\mathrm dt$ result in $e^{\frac{1}{2}\int_0^t\mathrm ds\lambda^2_s + \int_0^t\mathrm dW_s\lambda_s}$ as the full Jacobian up to time $t$.\\
Applying this transformation to the solution of the linear SME in the energy eigenbasis, one finds that the associated Jacobian factor is independent of the individual eigenvalues. Combined with the fact that the density matrix remains normalized at all times, this implies that the Jacobian is equal to unity,
\begin{eqnarray}
    \mathrm{Tr}[\rho(W_t)] =\mathrm{Tr}[\rho(\widetilde W_t)e^{\frac{1}{2}\int_0^t\mathrm ds\lambda^2_s + \int_0^t\mathrm dW_s\lambda_s}]=e^{\frac{1}{2}\int_0^t\mathrm ds\lambda^2_s + \int_0^t\mathrm dW_s\lambda_s}=1.
\end{eqnarray}
Consequently, the normalized nonlinear SME is statistically equivalent to the renormalized solution of the linear SME. In the present case, the nontrivial stochastic shift of the Wiener noise merely generates an alternative Wiener process, with observable differences arising only at the level of individual stochastic trajectories. \\

\begin{figure}
    \centering
    \includegraphics[width=1\columnwidth]{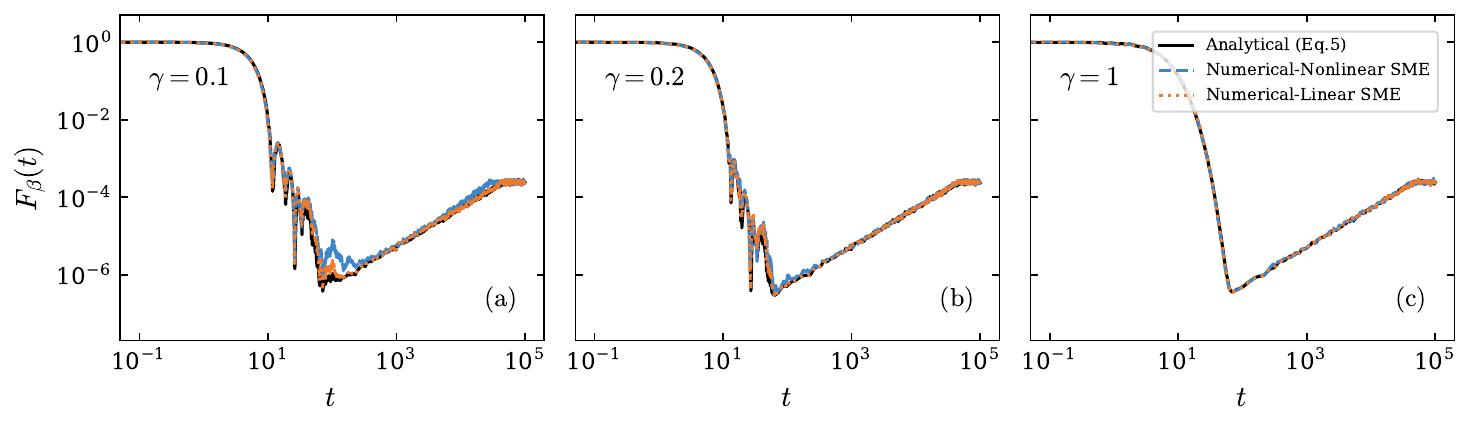}
    \caption{Evolution of SFF for different values of $\gamma$ in the SYK Hamiltonian with $N=26$. The figure illustrates the close correspondence between numerical simulations and analytical results.}
    \label{fig:Benchmarking}
\end{figure}

Figure ~\ref{fig:Benchmarking} shows the evolution of the SFF for different values of the dissipation strengths $\gamma$. The analytical curve corresponds to Eq.~(4) in the main text, which is the solution of the linear SME that was subsequently normalized. For larger $\gamma$, the numerical results from both linear and nonlinear SME closely follow the analytical solution, showing nearly perfect agreement. However, at lower $\gamma$, slight deviations in the nonlinear SSE evolution arise. Indeed, in the weak dissipation regime, the nonlinear terms --absent in the linear SME-- become comparable to the leading-order drift, leading to accumulated discrepancies over time.

\section{Stochastic Evolution of the average energy and its variance}
In order to investigate the thermodynamic consequences of stochastic dephasing, we investigate the evolution of energy and its variance under continuous monitoring. These quantities provide insight into how the energy distribution of the system fluctuates in time as a consequence of the noise realization. 
The instantaneous average energy $\langle H \rangle_t$ at a time $t$ and its variance $\langle \Delta H^2 \rangle_t$ can be expressed by means of equilibrium statistical physics, with the help of the ``dephased'' partition function, 
\begin{equation}
    Z(\beta-\sqrt{8\gamma}W_t,\gamma)=\sum_ne^{-(\beta -\sqrt{8\gamma}W_t)E_n-4\gamma E^2_nt},
\end{equation}
which generalizes the usual canonical partition function to include both deterministic dephasing and stochastic fluctuations. 

In terms of it, the average energy and its variance can be written as derivatives of its logarithm with respect to the inverse temperature $\beta$. 
\begin{eqnarray}
    \langle H\rangle_t&&=-\partial_\beta \log\left[Z(\beta-\sqrt{8\gamma}W_t,\gamma)\right]\\
    &&=\frac{\sum_nE_ne^{-(\beta -\sqrt{8\gamma}W_t )E_n-4t\gamma E^2_n}}{Z(\beta-\sqrt{8\gamma}W_t,\gamma)},\nonumber\\
    \langle \Delta H^2\rangle_t&&=\partial^2_\beta \log\left[Z(\beta-\sqrt{8\gamma}W_t,\gamma)\right].
\end{eqnarray}

These stochastic quantities satisfy the differential relation
\begin{eqnarray}
    \mathrm{d}\langle H\rangle_t &&= \mathrm{Tr}\left[ H \mathrm{d}\rho\right] \nonumber\\
    && = \sqrt{2\gamma}\mathrm{Tr}\left[H\left(\{H,\rho_t\}-2\langle H\rangle_t\rho_t\right)
\right]\mathrm{d}W_t \\
    &&=  2\sqrt{2\gamma}\left( \mathrm{Tr}\left[ H^2\rho \right] - (\mathrm{Tr} \left[H\rho\right])^2 \right)\mathrm{d}W_t\nonumber \\
    &&= \sqrt{8\gamma}\langle \Delta H^2\rangle_t\mathrm{d}W_t,
\end{eqnarray}
which connects fluctuations in the average energy to the instantaneous variance through the stochastic noise term. Clearly, the stochastic average $\mathbb{E}[\mathrm{d}\langle H\rangle_t]=0$. 
To gain further insight, one can consider the noise-average behavior of these quantities. For relatively short evolution times, the average can also be well captured by the annealed approximation, in which the averages of the numerator and denominator are taken independently over the Wiener process. Within this approximation, one finds
\begin{eqnarray}\label{eq: E_H_t}
    \mathbb E\left[\langle H\rangle_t\right]=\frac{\sum_n E_ne^{-\beta E_n}}{Z(\beta)},
\end{eqnarray}
which coincides with the equilibrium Boltzmann weight average at $t=0$. Thus, in the short-time or weak-dephasing limit, the stochastic dynamics preserves the equilibrium energy expectation value on average. \\

To see the behavior of the variance, let us introduce the conditional energy moments along a single stochastic trajectory, denoted by $M_{k,t}:=\langle H^k\rangle_t$, and the corresponding central moments, 
\begin{equation}
V_t:=\langle \Delta H^2\rangle_t=M_{2,t}-\mu_t^2,\qquad
\kappa_{3,t}:=\langle (H-\mu_t)^3\rangle_t=M_{3,t}-3\mu_tM_{2,t}+2\mu_t^3.
\end{equation}

From the stochastic master equation, one obtains
\begin{equation}
\mathrm{d}M_{2,t}=\mathrm{d}\langle H^2\rangle_t=\mathrm{Tr}(H^2\mathrm{d}\rho_t)=\sqrt{8\gamma}\big(M_{3,t}-\mu_t M_{2,t}\big)\mathrm{d}W_t,
\end{equation}

while $\mathrm{d}\mu_t = \sqrt{8\gamma}V_t d W_t$. Applying Itô's rule to $V_t$, 
\begin{equation}
    \mathrm{d}V_t = \mathrm{d}M_{2,t} - 2 \mu_t \mathrm{d}\mu_t - (\mathrm{d}\mu_t)^2,
\end{equation}
with $\mathrm{d}W_t^2 = \mathrm{d}t$, we get the stochastic differential equation for the closed variance, 

\begin{equation}
\mathrm{d}V_t=\sqrt{8\gamma} \kappa_{3,t}\mathrm{d}W_t-8\gamma V_t^2\mathrm{d}t.
\end{equation}

The variance is influenced by the instantaneous third centered moment $\kappa_{3,t}$. The Itô correction leads to a strictly negative drift proportional to $-V_t^2\mathrm{d}t$, which enforces a monotonic suppression of energy fluctuations along individual trajectories. Figure~\ref{suppfig:Variance} shows this behavior: for $\gamma=0$, the unitary evolution keeps the energy distribution stable, and the variance remains constant. However, for $\gamma>0$, the variance decays to zero, with increasing $\gamma$ accelerating the collapse. While individual trajectories show strong fluctuations at intermediate times, continuous monitoring eventually localizes the state within energy eigenbasis. The decay in variance observed here is a measurement-induced effect and does not occur in unmonitored dephasing dynamics, where the variance, $\langle\Delta H^2\rangle_t$ remains finite.

\begin{figure}[h]
    \centering
    \includegraphics[width=0.5\columnwidth]{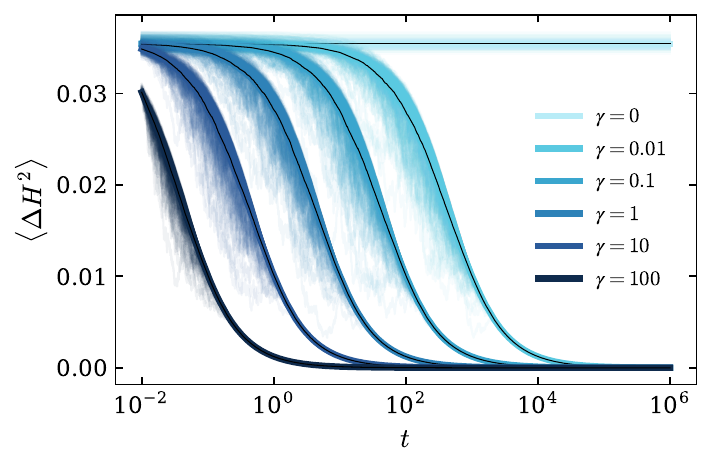}
    \caption{Time evolution of energy variance $\langle\Delta H^2\rangle_t$ under continuous monitoring for different measurement strengths $\gamma$. The light-colored curve represents individual stochastic trajectories, while the dark black line on it indicates the trajectory-averaged behavior, which is the average of 250 trajectory realizations.}
    \label{suppfig:Variance}
\end{figure}

\section{Self Averaging property of the SFF}

The SFF under dephasing Lindblad dynamics exhibits strong self-averaging, with individual realizations closely following the ensemble average; see Fig.~\ref{suppfig:SelfAveraging}(a). Dephasing in this setting leads to information loss, suppressing fluctuations around both the dip and plateau, so that individual trajectories show little deviation from the average~\cite{Xu21SFF}. Under continuous measurements, however, this self-averaging is lost as the measurement efficiency is increased. As shown in Fig.~\ref{suppfig:SelfAveraging}(b)-(c),  individual stochastic trajectories fluctuate significantly, reflecting the trajectory-dependent correlations induced by measurement back-action and highlighting the inherently stochastic nature of monitored quantum chaos.

\begin{figure}[h]
    \centering
    \includegraphics[width=1\columnwidth]{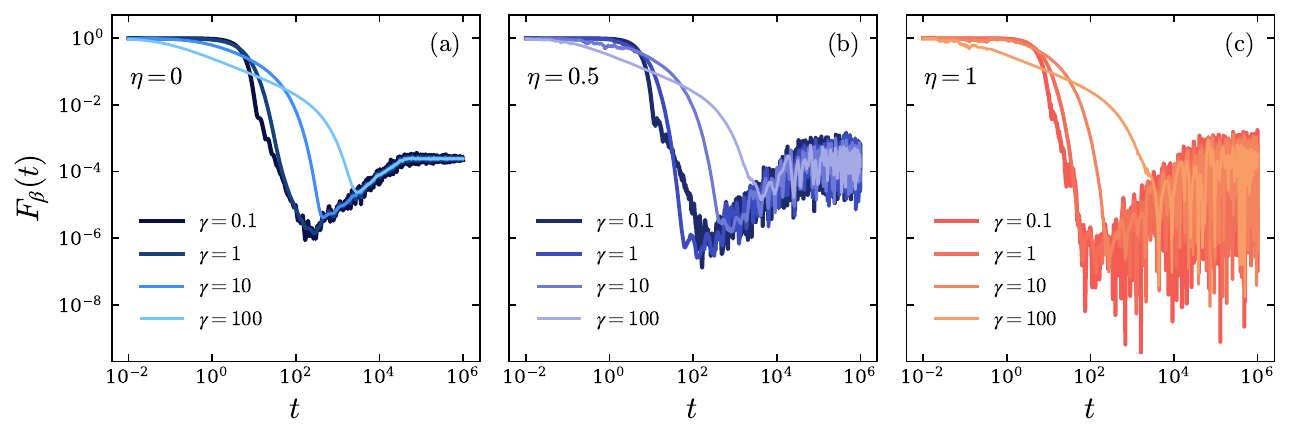}
    \caption{Evolution of SFF for different values of the measurement strength over a single SYK Hamiltonian ($N=26$) and a single noise trajectory. The case $\eta=0$ is associated with the mixed-state evolution resulting from pure dephasing in the absence of the innovation term. Fluctuations in time in the SFF are then suppressed. By contrast, at the level of single trajectories under continuous quantum measurements, the amplitude of the fluctuations increases with the value of $\eta$. The oscillatory behavior is maximal for $\eta=1$ when the conditioned quantum state is pure.}  
    \label{suppfig:SelfAveraging}
\end{figure}

\section{Universality of the monitored SFF across symmetry classes and models}

Here we present the averaged stochastic SFF for the quantum kicked top (QKT) and the Gaussian orthogonal ensemble (GOE). The QKT describes a spin $j$ subject to periodic rotations and nonlinear twists, with Floquet operator \cite{Haake1987,Ku1987,HaakeBook}
\begin{equation}
U_{0}=\exp\left(-i\frac{k}{2 j}J_z^2\right)\exp\left(-ipJ_y\right),
\end{equation}
where $J_{x,y,z}$ are collective angular-momentum operators, $p$ is the rotation angle, and $k$ is the twist strength. We set $p=\pi/2$. In the classical limit $j\to\infty$, the dynamics changes from regular to mixed near $k\approx 3$, and the phase space is predominantly chaotic for $k\gtrsim 6$ \cite{Haake1987,HaakeBook,Wang2004}. The model has been realized with cold atoms and superconducting qubits \cite{Chaudhury2009,Neill2016}. The parity $R_y=e^{-i\pi J_y}$ commutes with $U_0$. We therefore restrict to the even-parity sector, in which the quasienergy statistics follow the circular orthogonal ensemble (COE) \cite{Ku1987,HaakeBook}. We monitor the Floquet operator itself. Writing $U_0\equiv e^{-iH_F}$ with $H_F=\sum_n\theta_n|n\rangle\langle n|$ and $\theta_n\in(-\pi,\pi]$, continuous monitoring of $H_F$ is diagonal in the Floquet eigenbasis. It suppresses the coherence between $|n\rangle$ and $|m\rangle$ at a rate set by $\gamma(\theta_n-\theta_m)^2$.

By contrast, the GOE is a structureless ensemble whose eigenvalues exhibit Wigner--Dyson correlations. By the Bohigas--Giannoni--Schmit conjecture, these correlations are shared by generic quantum systems with chaotic classical limits \cite{BGS1984,Mehta2004}. We consider real symmetric $N\times N$ matrices $H$ drawn from the GOE,
\begin{equation}
P(H)\propto \exp\left(-\frac{N}{4\lambda^2}\mathrm{Tr},H^2\right),
\end{equation}
so that $\langle H_{ij}^2\rangle=\lambda^2/N$ for $i\neq j$ and $\langle H_{ii}^2\rangle=2\lambda^2/N$. The density of states is then a semicircle supported on $[-2\lambda,2\lambda]$. We set $\lambda=1$. The unitary SFF is known exactly for the GOE \cite{Brezin1997,Cotler2017,delCampo2017}. Both the COE and the GOE have Dyson index $\beta=1$ \cite{Dyson1962}, so the two models are compared at fixed symmetry class.

Both models display the phenomenology found for the SYK model in the main text. Beyond the post-selected no-jump evolution, typical trajectories show an enhanced ramp for weak monitoring [Fig.~\ref{suppFig:OtherModels}(a)--(c),(e)--(f)] and a suppressed ramp for strong monitoring [Fig.~\ref{suppFig:OtherModels}(d),(g)--(h)]. The monitoring-induced enhancement of the ramp, a spectral signature of the underlying level correlations, is therefore not specific to SYK. It originates from a spectral filter acting on the level correlations, as for decoherence and balanced gain and loss \cite{Xu2021,Cornelius2022}. The same qualitative non-monotonic dependence on $\gamma$ is observed in both the QKT and GOE, suggesting that the effect is not tied to the microscopic details of a particular model but instead reflects the interplay between continuous monitoring and Wigner--Dyson spectral correlations. The magnitude of the enhancement is model dependent and reflects the spectral properties of the underlying system. We leave its quantitative characterization to future work.

\begin{figure}
    \centering
    \includegraphics[width=1\columnwidth]{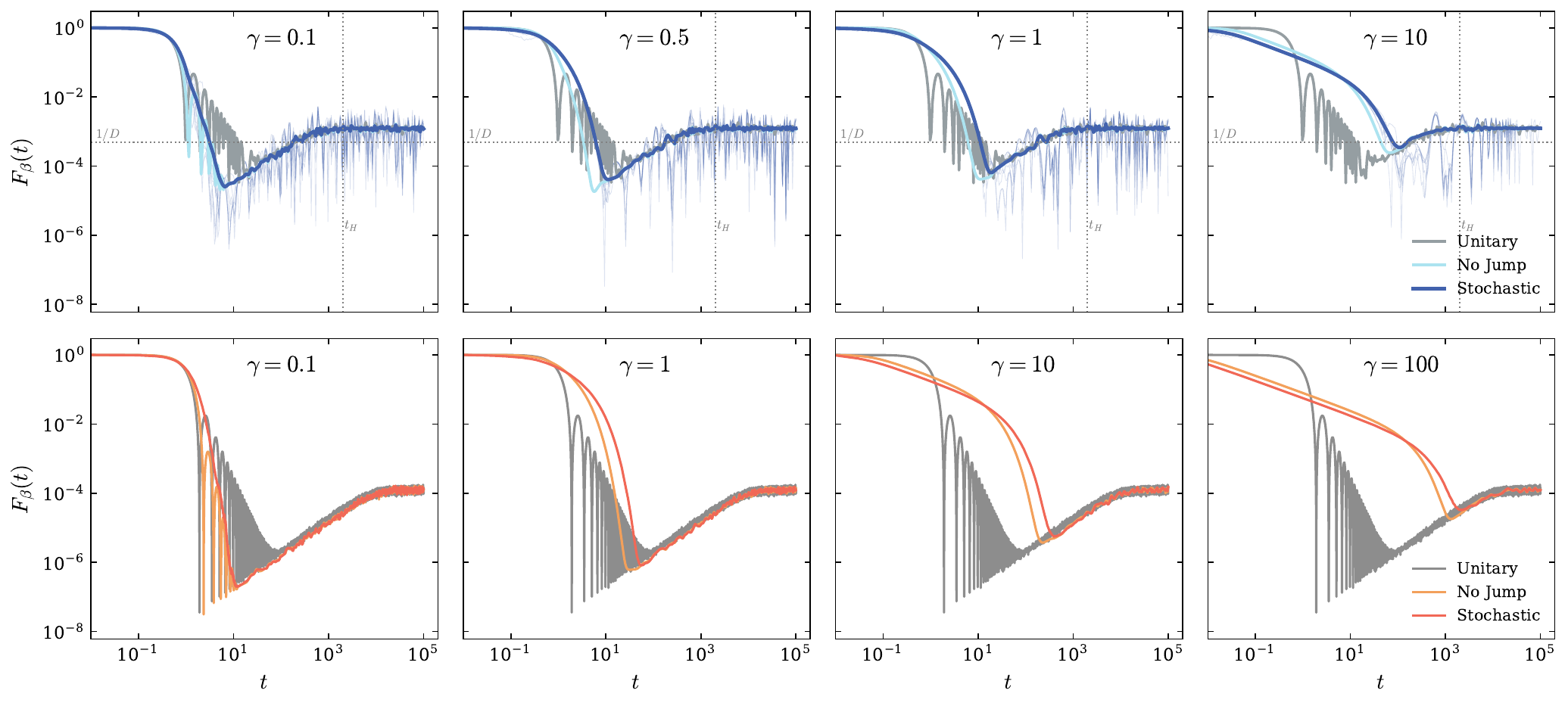}
    \caption{Evolution of spectral form factor under continuous monitoring for the Quantum Kicked Top and RMT for different measurement strengths, $\gamma$. Top Panel: monitored SFF for QKT of 800 spins, with the kicking strength averaged over $n_k=48$ values spaced in $k \in [9,11]$ to suppress non-universal fluctuations and over $60$ noise trajectories. Bottom Panel: monitored SFF for the GOE, $N=8000$  averaged over $200$ noise realizations. Individual stochastic realizations are shown in the background (light colored fluctuating trajectories). The SFF for unitary evolution is shown by the gray line. }
    \label{suppFig:OtherModels}
\end{figure}

\section{Finite Size Scaling}
Here we discuss the finite-size scaling of the three models considered. Fig.~\ref{Suppfig:Finitesize} shows $t_d/t_p$ as a function of the monitoring strength $\gamma$ for a range of system sizes $N$ (or Hilbert-space dimension $d$) for SYK, the QKT, and RMT. For all three models, the characteristic non-monotonic dependence on $\gamma$ persists across the system sizes considered: $t_d/t_p$ decreases with increasing $\gamma$ in the weak-monitoring regime, reaches a minimum at an intermediate monitoring strength, and subsequently increases in the strong-monitoring regime. Thus, the competition between the enhancement of spectral correlations at weak monitoring and their suppression by strong dephasing is already visible at finite size and does not rely on a particular system size.

As the system size is increased, the curves are shifted downward at fixed $\gamma$. This behavior is expected from the increasing plateau time $t_p$ with Hilbert-space dimension, while the characteristic deviation time $t_d$ remains controlled by the monitoring-induced modification of the spectral correlations. The overall shape of the $\gamma$ dependence, however, remains remarkably stable across system sizes. In particular, the position of the minimum changes only weakly with $N$, indicating that the characteristic monitoring strength at which the enhancement of the ramp is maximal is not strongly affected by finite-size effects. The persistence of both the weak- and strong-monitoring branches, together with the weak size dependence of the crossover scale, supports the interpretation of the non-monotonic response as a robust feature of the monitored dynamics rather than a finite-size artifact.

\begin{figure}[h]
    \centering
    \includegraphics[width=1\columnwidth]{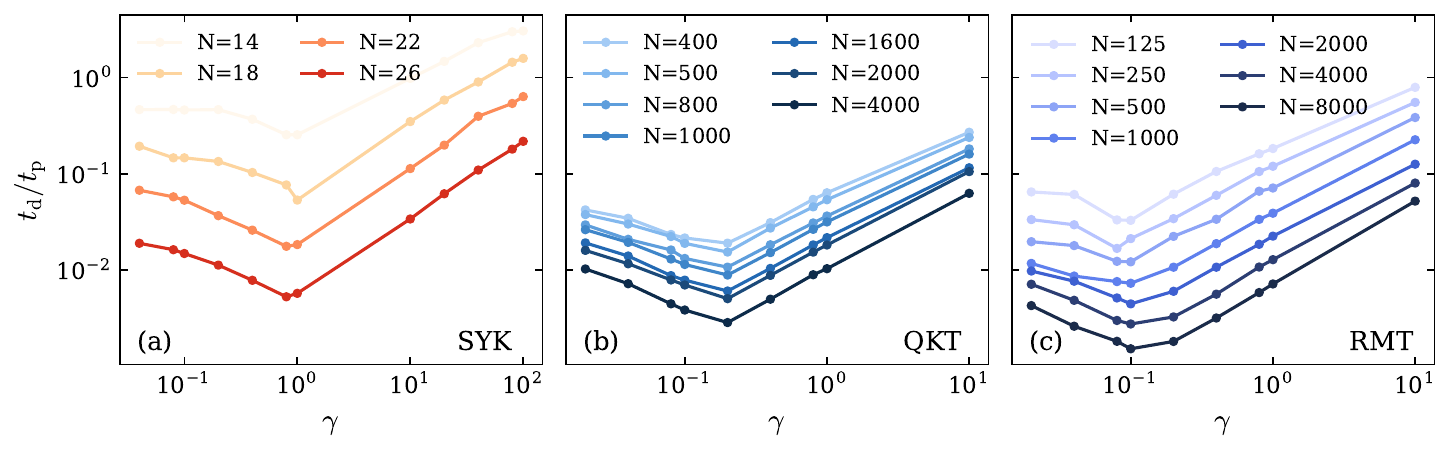}
    \caption{Dip-time ratio $t_d/t_p$ vs.\ $\gamma$ for (a) SYK, (b) QKT, and (c) RMT across system sizes $N$, averaged over $100$ noise and ensemble realizations.}

    \label{Suppfig:Finitesize}
\end{figure}

\section{Annealed approximation in Stochastic Averages}

Given  two functions $f(W_t)$ and $g(W_t)$ of the Wiener noise $W_t \sim \mathcal N(0,t)$, the stochastic average of their ratio can be approximated as
\begin{equation}
\mathbb{E}_{W_t}\left[\frac{f(W_t)}{g(W_t)}\right]=\frac{\mathbb{E}_{W_t}\left[f(W_t)\right]}{\mathbb{E}_{W_t}\left[g(W_t)\right]}-\frac{{\rm Cov}_{W_t}[f(W_t),g(W_t)]}{\mathbb{E}_{W_t}\left[g(W_t)\right]^2}+\frac{\mathbb{E}_{W_t}\left[f(W_t)\right]{\rm Var}_{W_t}[g(W_t)]}{\mathbb{E}_{W_t}\left[g(W_t)\right]^3}+\dots
\end{equation}
Thus, the annealed approximation holds whenever $|{\rm Cov}_{W_t}[f(W_t),g(W_t)]|\ll\mathbb{E}_{W_t}\left[g(W_t)\right]^2$ and ${\rm Var}_{W_t}[g(W_t)]\ll \mathbb{E}_{W_t}\left[g(W_t)\right]^2$.

Consider the annealed approximation for the stochastic average of SFF $F_\beta(t,W_t)$ with a fixed Hamiltonian. Provided that it holds,
using the identity
\begin{equation}
\mathbb E_{W_t}\left[
e^{\lambda W_t}
\right]
=
\exp\left(
\frac{\lambda^2 t}{2}
\right),
\end{equation}
the average of the numerator reads
\begin{eqnarray}
\mathbb{E}_{W_t}\left[\left|
\sum_{n}
e^{-(\beta+it-\sqrt{2\gamma}W_t)E_n}
\,e^{-2\gamma tE_n^2}
\right|^2\right]
&=&\mathbb{E}_{W_t}\left[\sum_{n,m}
e^{-(\beta+it)E_n}
e^{-(\beta-it)E_m}
e^{\sqrt{2\gamma}W_t(E_n+E_m)}
e^{-2\gamma t(E_n^2+E_m^2)}\right]\\
&=&\sum_{n,m}
e^{-(\beta+it)E_n}
e^{-(\beta-it)E_m}
\mathbb{E}\left[e^{\sqrt{2\gamma}W_t(E_n+E_m)}\right]
e^{-2\gamma t(E_n^2+E_m^2)}\\
&=&\sum_{n,m}
e^{-(\beta+it)E_n}
e^{-(\beta-it)E_m}
e^{\gamma t(E_n+E_m)^2}
e^{-2\gamma t(E_n^2+E_m^2)}\\
&=&\sum_{n,m}
e^{-(\beta+it)E_n}
e^{-(\beta-it)E_m}
e^{-\gamma t(E_n-E_m)^2}.
\end{eqnarray}
Similarly, the stochastic average of the denominator in the expression for the SFF yields
\begin{eqnarray}
\mathbb{E}_{W_t}\left[Z(\beta)Z(\beta-\sqrt{8\gamma}W_t,\gamma)\right]=\mathbb{E}_{W_t}\left[Z(\beta)\sum_ne^{-(\beta -\sqrt{8\gamma}W_t)E_n-4\gamma E^2_nt}\right]
=Z(\beta)^2,
\end{eqnarray}
and thus, in the annealed approximation
\begin{eqnarray}
\mathbb{E}_{W_t}\left[F_\beta(t,W_t)\right]\approx \frac{1}{Z(\beta)^2}\sum_{nm}e^{-\beta(E_n+E_m)-it(E_n-E_m)-\gamma t(E_n-E_m)^2},
\end{eqnarray}
which is the result under energy-dephasing dynamics, i.e., when $ \mathrm{d}\rho(t) = -i[H,\rho(t)] \mathrm{d}t - \gamma[H,[H,\rho(t)]] \mathrm{d}t$. 
From this, it follows that the breakdown of the annealed approximation is essential in our study to quantify the measurement-induced enhancement of quantum chaos under continuous quantum measurements relative to the case of energy dephasing. Physically, it reflects the measurement backaction, which is absent in the energy dephasing case.

In the study of spectral statistics, an additional average over an ensemble of Hamiltonians $\mathcal{E}(H)$ is also natural. This is the case in the SYK model under study due to the distribution of the couplings $J_{klmn}$. 
The study of the SFF involves then a second average $\mathbb{E}_{\mathcal{E}(H)}[\mathbb{E}_{W_t}\left[F_\beta(t,W_t)\right]]$. 
To analyze the role of the measurement backaction, we examine in Fig.~\ref{suppfig:Annealing_limit} the difference between the annealed and quenched definitions of the SFF under stochastic averaging. We consider three distinct averaging procedures for the stochastic SFF, 
\begin{equation}
F_\beta(t,W_t)=\frac{N(t,W_t)}{D(t,W_t)},
\end{equation}
expressed in terms of its numerator $N(t,W_t)$ and denominator $D(t,W_t)$. The first corresponds to an annealed average over Wiener noise for a fixed Hamiltonian,
\begin{equation}
\frac{\mathbb{E}_{W_t}\left[N(t,W_t)\right]}{\mathbb{E}_{W_t}\left[D(t,W_t)\right]}.
\end{equation}
The second performs annealed averages of the Wiener noise, and subsequently averages the ratio over Hamiltonian ensembles, 
\begin{equation}
    \mathbb{E}_{\mathcal{E}(H)}\left[\frac{\mathbb{E}_{W_t}\left[N(t,W_t)\right]}{\mathbb{E}_{W_t}\left[D(t,W_t)\right]}\right].
\end{equation}
Lastly, we consider the double use of the annealed approximation, both with respect to the Wiener noise and the Hamiltonian disorder,
\begin{equation}
\frac{\mathbb{E}_{\mathcal{E}(H)}[\mathbb{E}_{W_t}\left[N(t,W_t)\right]]}{\mathbb{E}_{\mathcal{E}(H)}[\mathbb{E}_{W_t}\left[D(t,W_t)\right]]}.
\end{equation}
In all cases, we evaluate the absolute value of relative error, $\Delta F_\beta(W_t,t)$, with respect to the fully quenched stochastic average,
\begin{equation}
\mathbb{E}_{\mathcal{E}(H)}\!\left[\mathbb{E}_{W_t}\!\left[F_\beta(t,W_t)\right]\right].
\end{equation}
\begin{figure}[h]
    \centering
    \includegraphics[width=1\columnwidth]{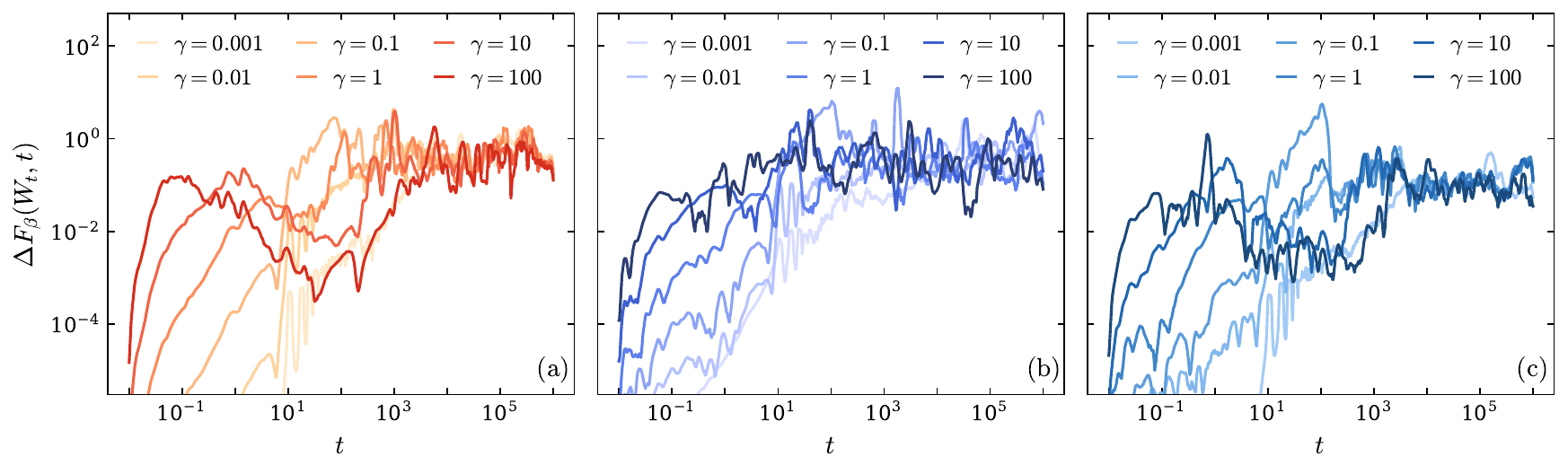}
    \caption{Time evolution of the relative error in the annealed average for the SFF, for different values of measurement strength $\gamma$. The panels correspond to (a) annealed averaging over Wiener noise realizations for a fixed Hamiltonian, (b) annealed averaging over Wiener noise followed by averaging over Hamiltonian ensemble realizations, and (c) annealed averaging over both Wiener noise and Hamiltonian ensemble realizations. The curves shown are moving averages of the raw data with a window size of 100. Data are averaged over 250 stochastic trajectories.}
    \label{suppfig:Annealing_limit}
\end{figure}
As Fig. \ref{suppfig:Annealing_limit} shows, the annealed approximation is effectively broken at all relevant time scales to diagnose quantum chaos by means of the SFF. In particular, it is pronounced during the dip, ramp, and plateau regimes, as manifested by the large values of $\Delta F_\beta(W_t,t)$ in the range of times $10^2-10^4$. The validity of the annealed approximation is restricted to short-time asymptotics when the SFF is close to unity and its early decay towards the dip.

\section{Inefficient measurements and Purity Decay}

In the framework of continuous quantum measurements, the system undergoes a stochastic yet pure evolution, such that the density matrix $\rho(t)$ remains pure at all times, i.e., $\mathrm{Tr}[\rho^2(t)] = 1$. This behavior is intuitive: in a perfectly efficient measurement, the measurement record contains the complete information extracted from the system, and the corresponding measurement backaction is fully correlated with this record. Consequently, the state of the system can be continuously updated based on observed measurement outcomes, yielding a single, well-defined pure-state trajectory that reflects the observer’s complete knowledge of the system. The stochastic fluctuations in the measurement record merely randomize the evolution of this pure state but do not induce decoherence, as no information is lost to unmonitored channels. However, averaging many such stochastic realizations effectively disregards the individual measurement outcomes, and the state of the system becomes mixed, reproducing the decoherence expected from the corresponding unconditional master equation.  

In contrast, when the measurement efficiency $\eta < 1$, the information acquired about the system is incomplete, and the corresponding quantum trajectories no longer remain pure. Consequently, purity decays as a function of both time and measurement efficiency $\eta$. 
The time-dependent density matrix for arbitrary $\eta$ reads
\begin{equation}
    \rho(t) = \frac{\sum_{nm}\rho_{nm}(0)e^{-i(E_n -E_m)t - \gamma t(E_n -E_m)^2 - \gamma \eta t (E_n + E_m)^2 +\sqrt{2\gamma\eta}W_t(E_n+E_m)}}{\sum_n \rho_{nn}(0)e^{-4\gamma \eta tE_n^2 + 2\sqrt{2\gamma\eta}W_tE_n}}\lvert n\rangle \langle m\rvert,
\end{equation}
whence it follows that
\begin{equation}\label{eq: P_t_W_t}
    P(t, W_t) =\mathrm{Tr}[\rho^2(t)]=\frac{\sum_{nm}\rho_{nm}(0)^2e^{-2\gamma t(E_n -E_m)^2 -2\gamma\eta t(E_n + E_m)^2 + 2\sqrt{2\gamma\eta}W_t(E_n+E_m)}}{(\sum_n \rho_{nn}(0)e^{-4\gamma \eta tE_n^2 + 2\sqrt{2\gamma\eta}W_tE_n})^2}.
\end{equation}
Alternatively, using the Hubbard-Stratonovich transformation,
\begin{eqnarray}
e^{-a E^2 }
=\frac{1}{\sqrt{4\pi a}}\int{\rm d}ye^{-\frac{y^2}{4a}-iy E}, \quad {\rm Re}(a) >0,
\end{eqnarray}
and the fact that for the initial coherent Gibbs state $\rho_{nm}(0)=e^{-\beta(E_n+E_m)/2}/Z(\beta)$, the purity in Eq.~\eqref{eq: P_t_W_t} can be expressed as
\begin{eqnarray}
P(t,W_t)&=&\frac{\int\mathrm dy\int\mathrm dz e^{-\frac{z^2}{8\gamma\eta t}}e^{-\frac{y^2}{8\gamma t}}\sum_{nm}\rho^2_{nm}(0)e^{-iy(E_n-E_m)}e^{-i\left(z-2i\sqrt{2\gamma\eta}W_t\right)(E_n+E_m)}}{\left(\int \mathrm dye^{-\frac{y^2}{16\gamma\eta t}}\sum_n\rho_{nn}(0)e^{-i\left(y+2i\sqrt{2\gamma\eta}W_t\right)E_n}\right)^2}\\
&=&\frac{\int\mathrm dy\int\mathrm dz e^{-\frac{z^2}{8\gamma\eta t}}e^{-\frac{z^2}{8\gamma t}}Z\left(\beta-2\sqrt{2\gamma\eta W_t}+i(y+z)\right)Z\left(\beta +i(z-y)\right)}{\left(\int\mathrm dy\,e^{-\frac{y^2}{16\gamma\eta t}}Z(\beta+iy-2\sqrt{2\gamma\eta}W_t)\right)^2}.\nonumber
\end{eqnarray}
These expressions explicitly capture the interplay between measurement inefficiency and dephasing, providing a quantitative description of purity decay in continuously monitored quantum systems.

It is insightful to consider the stochastic average of the purity.
In the annealed approximation,
\begin{equation}
\mathbb{E}_{W_t}[P(t,W_t)]\approx \frac{\sum_{nm}
\rho_{nm}(0)^2
e^{-2\gamma t(E_n -E_m)^2 + 2\gamma \eta t(E_n + E_m)^2}}
{\sum_{nm}
\rho_{nn}(0)\rho_{mm}(0)
e^{-8\gamma \eta tE_nE_m}}.
\end{equation}
By contrast, under energy dephasing, using the annealed approximation for $\mathbb{E}[\rho(t)]$, that reads as
\begin{equation}
\mathbb{E}_{W_t}\left[\rho(W_t,t)\right]\approx 
\sum_{nm}\rho_{nm}(0)
 e^{-it(E_n-E_m)-\gamma t(E_n-E_m)^2}\lvert n\rangle \langle m\rvert.
\end{equation}
the purity is given by
\begin{equation}
P(t)=\mathrm{Tr}\!\left[\mathbb{E}_{W_t}\!\left[\rho(W_t,t)\right]^2\right]
\approx \sum_{nm}\rho_{nm}(0)^2\,e^{-2\gamma t(E_n-E_m)^2}.
\end{equation}

\section{Enhancement of Chaos}

From Fig.~1 in the main text, we observe that, as the measurement strength $\gamma$ increases, the stochastic SFF crosses over from the no-jump limit to behavior characteristic of energy-dephasing dynamics.
This behavior can be understood in terms of the decoherence induced by energy dephasing.
The off-diagonal contributions to the coherence of the SFF decay over a timescale given by ~\cite{Xu19,Xu21SFF}
\begin{equation}
    t_{\rm dec} =\frac{1}{4\gamma\frac{\mathrm d^2}{\mathrm d\beta^2}Z(\beta)}=
    \frac{1}{4\gamma\mathrm{Var}_\beta(H)},
\end{equation} 
where $\mathrm{Var}_\beta(H)$ denotes the equilibrium energy variance in the Gibbs state at inverse temperature $\beta$.

This sets the timescale beyond which energy-difference–dependent dephasing becomes significant in the stochastic dynamics.\\
At the same time, off-diagonal contributions to the SFF are restricted by the Gaussian factor $e^{-2\gamma t E_n^2}$ for each trajectory as
\begin{equation}
    2\gamma t E^2 \lesssim 1, \implies |E| \lesssim \frac{1}{\sqrt{\gamma t}}.
\end{equation}
 The effects of quantum jumps are captured by the Wiener-noise factors $e^{\sqrt{2\gamma} W_t E_n}$, which, on average, decay on the same characteristic timescale and therefore do not modify the overall spectral filtering. More precisely, for energies $|E|\lesssim 1/\sqrt{\gamma t}$, one also has $\sqrt{\gamma} W_t E \lesssim 1$, since typical realizations of the Wiener process scale with its standard deviation, $W_t\sim\sqrt t$.
In the weak monitoring regime, $\gamma\ll 1$, where $t_\mathrm{d}\sim\gamma^{-1/2}$, it follows that $t_\mathrm{dec}\gg t_\mathrm{d}$. As a result, coherences affected by energy dephasing lie deep inside the filtered spectral regions,
\begin{eqnarray}
|E|\sim\frac{1}{\sqrt{\gamma t_\mathrm{dec}}}\ll\frac{1}{\sqrt{\gamma t_\mathrm{d}}}.
\end{eqnarray}
In other words, the short-range energy correlations responsible for coherence remain essentially unaffected by the Gaussian filter. The filter predominantly suppresses long-range correlations and spectral-edge contributions, thereby enhancing chaotic signatures.

For strong monitoring, $\gamma\gg1$, an opposite scenario traces out.While $t_\mathrm{dec}$ continues to decrease with increasing $\gamma$, the dip time grows linearly, $t_\mathrm{d}\sim\gamma$, exceeding the decoherence time, $t_\mathrm{d}\gg t_\mathrm{dec}$.
The Gaussian filters start suppressing the very same short-range correlations as in energy dephasing dynamics. Accordingly, for $\gamma\gg 1$, the SFF converges to that of pure energy dephasing dynamics. 
These considerations quantitatively characterize the crossover from coherent spectral dynamics to energy-dephasing–dominated behavior, controlled jointly by the measurement strength $\gamma$ and evolution time through the scaling variable $\gamma t$.

\section{Spectral Form Factor under varying measurement efficiency}

\begin{figure}[h]
    \centering
    \includegraphics[width=0.45\columnwidth]{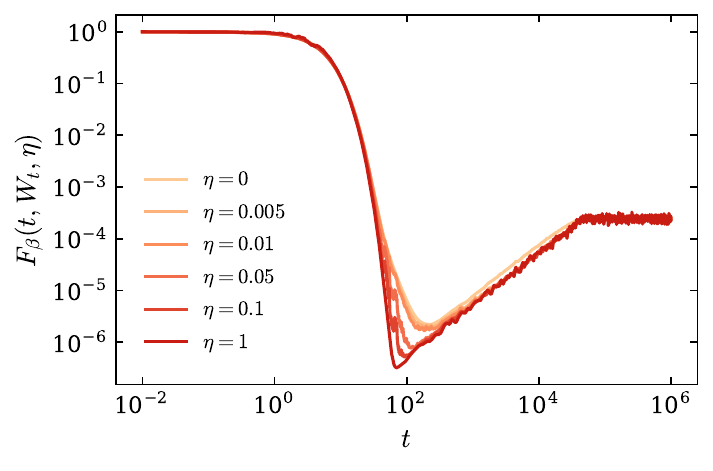}
    \caption{Evolution of SFF under different measurement efficiency for a fixed measurement strength $\gamma =1$, averaged over $250$ stochastic realisations.}
    \label{suppfig:SFF_ineff}
\end{figure}

As depicted in Fig.~\ref{suppfig:SFF_ineff}, for all efficiencies, the SFF displays the characteristic dip–ramp–plateau structure associated with chaotic dynamics. At early times, the curves collapse, indicating that short-time dynamics is largely insensitive to the measurement efficiency. However, at intermediate times, with an increase in $\eta$, the dip becomes progressively deeper and sharper. The sharpening effects can be understood in terms of enhanced information extraction and higher measurement efficiency, which purify individual trajectories and strengthen spectral correlations. Enhanced quantum chaos leads to a deeper dip in the SFF and an earlier onset of the ramp, which is sharper and longer.

\section{STOCHASTIC SPECTRAL FORM FACTOR AND DIP TIME}

To analyze the effect of continuous measurements on spectral statistics, it is useful to recall that the SFF can be decomposed into three contributions, each carrying different spectral information. Writing $\lvert \sum_n A_n \rvert^2 = \sum_{n,m} A_n A_m^*$, one finds a sum of a diagonal term with $n=m$, and an off-diagonal contribution with $n \neq m$. The latter further splits into a disconnected component without correlations among different eigenvalues and a connected component that encodes genuine two-level correlations. This decomposition reads 
\begin{eqnarray}
    F_\beta(t,W_t) = F_\beta^{(diag)}(t,W_t) + F_\beta^{(disc)}(t,W_t) + F_\beta^{(conn)}(t,W_t),
\end{eqnarray}
where $F^{(diag)}$ is time-independent and determines the contribution of self-correlation (plateau), $F^{(disc)}$ is governed by the density of states, and $F^{(conn)}$ reflects universal level-repulsion. 
The energy ensemble average of the SFF in the annealing limit can be written as  
\begin{align}
\label{eq:SFF_average}
\mathbb{E}_{\mathcal{E}(H)}[F_\beta(t,W_t)]  &= \frac{\mathbb{E}_{\mathcal{E}(H)}\left[\left|\sum_n A_n(E_n,W_t) \right|^2\right]}{\mathbb{E}_{\mathcal{E}(H)}\left[Z(\beta)\, Z(\beta-\sqrt{8\gamma} W_t, \gamma)\right]},
\end{align}
where we have defined $A_n(E_n,W_t) = e^{-(\beta+ i t) E_n} \, e^{-2\gamma t E_n^2 + \sqrt{2\gamma} W_t E_n}.$ In terms of density of states, 
\begin{eqnarray}
\mathbb{E}_{\mathcal{E}(H)}[F_\beta(t,W_t)]= \frac{\iint \mathrm{d}E ~ \mathrm{d}E' \mathbb{E}_{\mathcal{E}(H)}[\varrho(E,E')]A(E,W_t)A^*(E',W_t)}{\mathbb{E}_{\mathcal{E}(H)}\left[Z(\beta)\, Z(\beta-\sqrt{8\gamma} W_t, \gamma)\right]},
\end{eqnarray}
where $\varrho(E,E')$ is the two-point spectral density. Splitting 
\begin{equation}
\mathbb{E}_{\mathcal{E}(H)}[\varrho(E,E')] =  \mathbb{E}_{\mathcal{E}(H)}[\varrho(E)]\delta(E -E') + \mathbb{E}_{\mathcal{E}(H)}[\varrho(E)\ \varrho(E')] + \mathbb{E}_{\mathcal{E}(H)}[\varrho_c(E,E')],
\end{equation}
makes the diagonal, disconnected, and connected contributions explicit. \\

For a large $N$, the DOS of SYK is given by 
\begin{eqnarray}
     \mathbb{E}_{\rm SYK}[ \varrho(E)]&& = \frac{1}{2\pi}\int {\rm d}t e^{-iEt}\mathbb{E}_{\rm SYK}[ e^{iHt}]\\
     && \simeq \sqrt{\frac{2}{\pi N}}d\,\, \mathrm{exp} \left( -\frac{2E^2}{N} \right).
     \end{eqnarray} 
   
The ensemble average of the partition function and the dephased partition function are

\begin{eqnarray}
    \mathbb{E}_{\rm SYK}[Z(x)] \ && \simeq d \ \mathrm{exp} \left( \frac{Nx^2}{8}\right), \\
  \mathbb{E}_{\rm SYK}[Z(x, \gamma)]\ && \simeq d \left( \frac{1}{\sqrt{1 + 2N\gamma t}}\right) \ \mathrm{exp}\left( \frac{Nx^2}{8 + 16N\gamma t } \right).
\end{eqnarray}

The connected part of the two-point correlation function of GUE (i.e., $N \ \mathrm{mod} \ 8 = 2 \ \mathrm{or} \ 6$ ) takes the form 
 \begin{align}
     \mathbb{E}_{\rm SYK}[ \varrho_c(E,E')] \simeq - \left(\frac{\sin\left(2\pi r  \mathbb{E}_{\rm SYK}[\varrho(\omega)]\right)}{\pi r}\right)^2,
 \end{align}
with $r=E-E'$ and $\omega = (E+E')/2$. 
The diagonal part of the SFF then reads 
\begin{eqnarray}
\label{eq:SFF_Fdiag}
    \mathbb{E}_{\rm SYK}[F_\beta^{(diag)}(t\rightarrow\infty,W_t)] =  \frac{2\mathbb{E}_{\rm SYK}[ Z(2\beta - \sqrt{8\gamma}W_t,\gamma)]}{\mathbb{E}_{\rm SYK}[ Z(\beta)] \mathbb{E}_{\rm SYK}[ Z(\beta - \sqrt{8\gamma}W_t,\gamma)]} \simeq \frac{2}{d} e^{-N \beta^2 /8}.
\end{eqnarray}
The disconnected term captures the coarse features of DOS and governs the early-time behavior of the SFF prior to the onset of correlations. Using the Gaussian form of the SYK density of states, the disconnected contribution takes the form
\begin{eqnarray}
 \label{eq:SFF_Fdisc}
 \mathbb{E}_{\rm SYK}[F_\beta^{\rm (disc)}(t,W_t)] \ &&= \frac{\mathbb{E}_{\rm SYK}[\left|\int {\rm d}E\varrho(E) e^{-(\beta - \sqrt{2\gamma} W_t + i t) E_n} \, e^{-2\gamma t E_n^2} \right|^2]}{\mathbb{E}_{\rm SYK}[Z(\beta)]\, \mathbb{E}_{\rm SYK}[Z(\beta-\sqrt{8 \gamma} W_t, \gamma)]} \nonumber \\
 && = \frac{\frac{1}{1 + \gamma N t} \Big|\exp \left[\frac{N(\beta + i t - \sqrt{2\gamma}W_t)^2}{8 + 8\gamma N t} \right] \Big| ^2}{\frac{1}{\sqrt{1 + 2\gamma N t}} \exp \left[ \frac{N \beta ^2}{8}\right] \exp \left[ \frac{N(\beta-\sqrt{8\gamma}W_t)^2}{8 + 16N\gamma t }\right] },
\end{eqnarray}

while the connected part is given by
\begin{eqnarray}
\label{}
\mathbb{E}_{\rm SYK}[F_\beta^{(\mathrm{conn})}(t, W_t)]
= -\,\frac{
\displaystyle \int_{-\infty}^{\infty} \!\mathrm{d}\omega 
\int_{-\infty}^{\infty} \! \mathrm{d}r\;
\left(\frac{\sin\!\left(2\pi r\, \mathbb{E}_{\rm SYK}[ \varrho(\omega)] \right)}{\pi r}\right)^{\!2}
\exp\!\Big[
-2\beta \omega - i t r - \gamma t r^2 - 4\gamma t \omega^2
+ 2\sqrt{2\gamma}\,W_t\, \omega
\Big]}{\mathbb{E}_{\rm SYK}[Z(\beta)]\, \mathbb{E}_{\rm SYK}[Z\!\left(\beta - \sqrt{8\gamma}\, W_t, \gamma\right)]}.\nonumber\\
\end{eqnarray}

The integral over the relative coordinate $r$ can be evaluated in a closed form. In the strong dissipation regime $(\gamma \gg 1)$, for late times $t \gg d/\sqrt{N}$, where the Gaussian suppresses all but the narrow support of the sinc kernel, one obtains
\begin{eqnarray}
    &&\int_{-\infty}^\infty\,\mathrm dr \left(\frac{\sin(k(\omega)r)}{\pi r}\right)^2 e^{-itr-\gamma tr^2}
   \approx\frac{k(\omega)}{\pi},
\end{eqnarray}
with $k(\omega)=2\pi \mathbb{E}_{\rm SYK}[\varrho(\omega)]=2\pi \sqrt{\frac{2}{\pi N}}d\,e^{-\frac{2\omega^2}{N}}$. Carrying out the remaining $\omega$-integral and normalizing by the dephased partition function yields, at leading order in $t$, 
\begin{equation}
\label{eq:SFF_Fconn}
    \mathbb{E}_{\rm SYK}[F^{(conn)}_{\beta} (t,W_t)] \simeq \frac{1}{2d^2} \sqrt{\frac{N}{8\pi}}~t \exp \left[ \frac{\beta^2}{4 t \gamma} -\frac{\beta W_t}{\sqrt{2t\gamma}} + \frac{W_t^2}{4 N \gamma t^2} -\frac{N\beta^2}{8}\right].
\end{equation}
The dip time $t_{\mathrm d}$ is determined by crossing the disconnected and connected pieces. Using $F_\beta^{(disc)}(t,W_t = 0) = F_\beta ^{(conn)}(t,W_t=0)$ at $\beta=0$, we get the dip time,  

\begin{eqnarray}
    t_{\mathrm d}(\gamma) = 6 \gamma~W\left( \frac{1}{6\gamma}\left(\frac{8 d ^2 \sqrt{2\pi}}{n\sqrt{\gamma}}\right)^{2/3}\right),
\end{eqnarray}
where $W(x)$ is the Lambert function.
Note that Eqs. \eqref{eq:SFF_Fdiag}, \eqref{eq:SFF_Fdisc} and \eqref{eq:SFF_Fconn} hold only for small $\beta$. However, they can also be used to estimate the SFF for finite $\beta$.

\section{Out-of-time-order correlators under continuous monitoring}

We next provide further evidence on how quantum chaos is tailored by continuous
monitoring. To this end,  we study the out-of-time-ordered correlator (OTOC) in the SYK,
RMT, and kicked-top models. We use the coherent Gibbs state
\[
\ket{\psi_\beta}=Z(\beta)^{-1/2}\sum_n e^{-\beta E_n/2}\ket{n},
\]
where $H=\sum_n E_n\ket{n}\!\bra{n}$ and $Z(\beta)=\sum_n e^{-\beta E_n}$.
We normalize the Majorana operators as $\chi_i^2=\mathbb I$, so that
$\{\chi_i,\chi_j\}=2\delta_{ij}$.\\

For a given realization of the measurement noise $W_t$, the monitored state
acquires an energy-dependent Gaussian filter. Its normalization is
\begin{equation}
\mathcal N(t,W_t)
=\frac{1}{Z(\beta)}\sum_k
e^{-\beta E_k-4\gamma tE_k^2+2\sqrt{2\gamma}W_tE_k}
=\frac{Z(\beta-\sqrt{8\gamma}W_t,\gamma)}{Z(\beta)}.
\end{equation}
The corresponding unnormalized Heisenberg evolution of an operator $O$ is
\begin{equation}
\widetilde O_{nm}(t,W_t)
=O_{nm}(0)e^{i(E_n-E_m)t-2\gamma t(E_n^2+E_m^2)
+\sqrt{2\gamma}W_t(E_n+E_m)}.
\end{equation}
For a Majorana operator $\chi_i$, this gives
\begin{equation}\label{Eq:SM_chitilde}
\widetilde{(\chi_i)}_{nm}(t,W_t)
=(\chi_i)_{nm}
e^{i(E_n-E_m)t-2\gamma t(E_n^2+E_m^2)
+\sqrt{2\gamma}W_t(E_n+E_m)}.
\end{equation}
Thus, in addition to the usual unitary phase evolution, continuous monitoring
introduces a Gaussian energy filter. In the no-jump limit $W_t=0$, this filter
gives the deterministic envelope
\[
\mathcal N_{\rm nj}(t)
=\frac{1}{Z(\beta)}
\sum_k e^{-\beta E_k-4\gamma tE_k^2},
\]
while $\gamma=0$ recovers the unitary dynamics. Note that the real part of
the exponent in Eq.~\eqref{Eq:SM_chitilde} is symmetric in $n$ and $m$, so
$\widetilde\chi_i(t,W_t)$ is Hermitian only at $\gamma=0$; at finite $\gamma$
it is not, and this will matter in the following.\\

We characterize scrambling using the normalized four-point function
\begin{equation}
\mathcal F_\beta(t,W_t)
=\frac{F_\beta(t,W_t)}{F_\beta(0)},
\end{equation}
with
\begin{equation}
F_\beta(t,W_t)
=\frac{1}{\mathcal N(t,W_t)}
\bra{\psi_\beta}
\widetilde\chi_i(t,W_t)\chi_j
\widetilde\chi_i(t,W_t)\chi_j
\ket{\psi_\beta}.
\end{equation}
By construction, $\mathcal F_\beta(0)=1$, and $F_\beta(0)=-1$, since
anticommuting the two Majoranas past one another and using
$\chi_i^2=\chi_j^2=\mathbb I$ produces a single sign. In the unitary problem, the decay of
this quantity signals the spreading of the initially localized operator.
For anticommuting Majoranas, the corresponding diagnostic is the squared
anticommutator,
\begin{equation}
C_{ij,\beta}(t,W_t)
=\frac{1}{\mathcal N}
\left\langle
\{\widetilde\chi_i(t,W_t),\chi_j\}^{\dagger}
\{\widetilde\chi_i(t,W_t),\chi_j\}
\right\rangle_{\psi_\beta}.
\end{equation}
Writing $\widetilde W\equiv\widetilde\chi_i(t,W_t)$ and $V\equiv\chi_j$ and
expanding the product,
\begin{equation}\label{Eq:SM_expand}
\{\widetilde W,V\}^\dagger\{\widetilde W,V\}
=\underbrace{V\widetilde W^\dagger\widetilde WV
+\widetilde W^\dagger V^2\widetilde W}_{\text{time-ordered}}
+\underbrace{V\widetilde W^\dagger V\widetilde W
+\widetilde W^\dagger V\widetilde WV}_{\text{out-of-time-order}} ,
\end{equation}
which defines the time-ordered contribution
\begin{equation}\label{Eq:SM_Gdef}
G_\beta(t,W_t)
=\frac{1}{2\mathcal N}
\langle\psi_\beta\lvert
\chi_j\widetilde\chi_i^\dagger\widetilde\chi_i\chi_j
+\widetilde\chi_i^\dagger\chi_j^2\widetilde\chi_i
\rvert\psi_\beta\rangle .
\end{equation}
The filter is symmetric in $n$ and $m$ and real, so $\widetilde\chi_i$
remains Hermitian at any $\gamma$, with $\widetilde\chi_i^\dagger=\widetilde
\chi_i$. The two out-of-time-order terms in Eq.~\eqref{Eq:SM_expand} are
therefore mutual Hermitian conjugates and sum to $2\Re F_\beta$, while the
time-ordered terms reduce, using $\chi_j^2=\mathbb I$, to
$\widetilde\chi_i^2+\chi_j\widetilde\chi_i^2\chi_j$. It follows that
\begin{equation}
C_{ij,\beta}(t)
=2\left[G_\beta(t)+\Re F_\beta(t)\right] ,
\end{equation}
at any monitoring strength, and with $F_\beta(0)=-1$,
\begin{equation}
\mathcal F_\beta(t)
=1-\frac{C_{ij,\beta}(t)}{2|F_\beta(0)|}.
\end{equation}
In the unitary limit $\widetilde\chi_i^2=\mathbb I$, so $G_\beta=1$ and the decay of the OTOC is exactly equivalent to the growth of the squared anticommutator. At finite $\gamma$, the energy filter damps the matrix elements non-uniformly, $\widetilde\chi_i^2\neq\mathbb I$ and $G_\beta$ drops below unity, so the two diagnostics differ by that amount. We work directly with $\mathcal F_\beta$ throughout, which is the quantity computed numerically, and use the squared anticommutator only to fix its short-time form.\\

The short-time behavior is different from the later chaotic growth. In the
unitary case, expanding $\widetilde\chi_i(t)=\chi_i+it[H,\chi_i]+\mathcal
O(t^2)$ and using $\{\chi_i,\chi_j\}=0$ for $i\neq j$, the anticommutator starts
at order $t$,
\begin{equation}
\{\widetilde\chi_i(t),\chi_j\}=t\big\{B_i,\chi_j\big\}+\mathcal O(t^2),
\qquad B_i\equiv i[H,\chi_i]=B_i^\dagger ,
\end{equation}
so that
\begin{equation}
C_{ij,\beta}(t)=A t^2+\cdots,
\qquad
A=\langle\psi_\beta\lvert\{B_i,\chi_j\}^2\rvert\psi_\beta\rangle\ge0 ,
\end{equation}
the positivity being manifest because $\{B_i,\chi_j\}$ is Hermitian. Hence
\begin{equation}
\mathcal F_\beta(t)
=1-\frac{A}{2|F_\beta(0)|}t^2+\mathcal O(t^4).
\end{equation}
The initial decay is therefore quadratic, with a vanishing linear slope. This behavior is fixed by the short-time dynamics and does not by itself characterize quantum chaos.\\ 

Continuous monitoring changes this behavior immediately. Each matrix element of the four-point function acquires the Gaussian factor
\begin{equation}
e^{-2\gamma t(E_a^2+E_b^2+E_c^2+E_d^2)},   
\end{equation}
that is, $e^{-2\gamma tE^2}$ for each of the four energy labels entering the four-point function. Replacing each by its typical value, the numerator is damped as $e^{-8\gamma t\langle E^2\rangle_\beta}$, while the normalization $\mathcal N$ carries a single filter $e^{-4\gamma tE_k^2}$ and is damped as $e^{-4\gamma t\langle E^2\rangle_\beta}$. Dividing leaves the net rate
\begin{equation}\label{Eq:SM_Gamma}
\Gamma\simeq4\gamma\langle E^2\rangle_\beta ,
\end{equation}
which at $\beta=0$ equals $4\gamma\,\mathrm{Var}_\beta(E)$ since $\langle E\rangle_\beta=0$. Even at weak monitoring the OTOC therefore acquires an extra damping whose rate grows linearly with $\gamma$. This filtering effect is distinct from scrambling itself, and it does not cancel. The normalization removes only half of the numerator's damping, in contrast to the SFF, where the envelope cancels between the fidelity and its trajectory normalization and leaves the ramp free to be extended. 

At later times, the intrinsic chaotic dynamics becomes visible. In the chaotic regime, the squared anticommutator grows approximately as
\begin{equation}
C_{ij,\beta}(t)\sim N^{-1}e^{\lambda_L t},
\end{equation}
leading to
\begin{equation}
\mathcal F_\beta(t)
\simeq1-\frac{c}{N}e^{\lambda_L t}.
\end{equation}
We extract the growth exponent from the log-linear regime of $\mathcal F_\beta(t)$ in the intermediate-time window, before the first minimum of the OTOC, and denote it $\lambda_{\rm eff}$ to keep it distinct from the intrinsic $\lambda_L$. The effect of monitoring in this regime can be understood by separating the measurement-induced decay from the intrinsic operator growth. To leading order,
\begin{equation}
\mathcal F_\beta(t)
\approx e^{-\Gamma t}\mathcal F_\beta^{(0)}(t),
\end{equation}
with $\Gamma$ given by Eq.~\eqref{Eq:SM_Gamma} and
$\mathcal F^{(0)}_\beta$ the unmonitored correlator, so that for
$\Gamma t\ll1$
\begin{equation}\label{Eq:SM_OTOCdecay}
1-\mathcal F_\beta(t)
\approx
\Gamma t+\frac{c}{N}e^{\lambda_L t}.
\end{equation}
The first term is produced by the measurement filter, while the second describes chaotic operator growth. Consequently $\lambda_{\rm eff}$ exceeds $\lambda_L$ and grows with $\gamma$, as seen in Fig.~\ref{fig:otocgrid}, where it remains at $\simeq0.56\,J$ up to $\gamma\simeq0.05$ before rising. This steepening should be distinguished from a genuine change of the underlying chaotic exponent: at weak monitoring it is entirely a consequence of the additional filter-induced decay. A genuine modification of the chaotic dynamics may nevertheless arise because the filter progressively concentrates the surviving operator weight in the central spectral region.

\begin{figure}
    \centering
    \includegraphics[width=1\columnwidth]{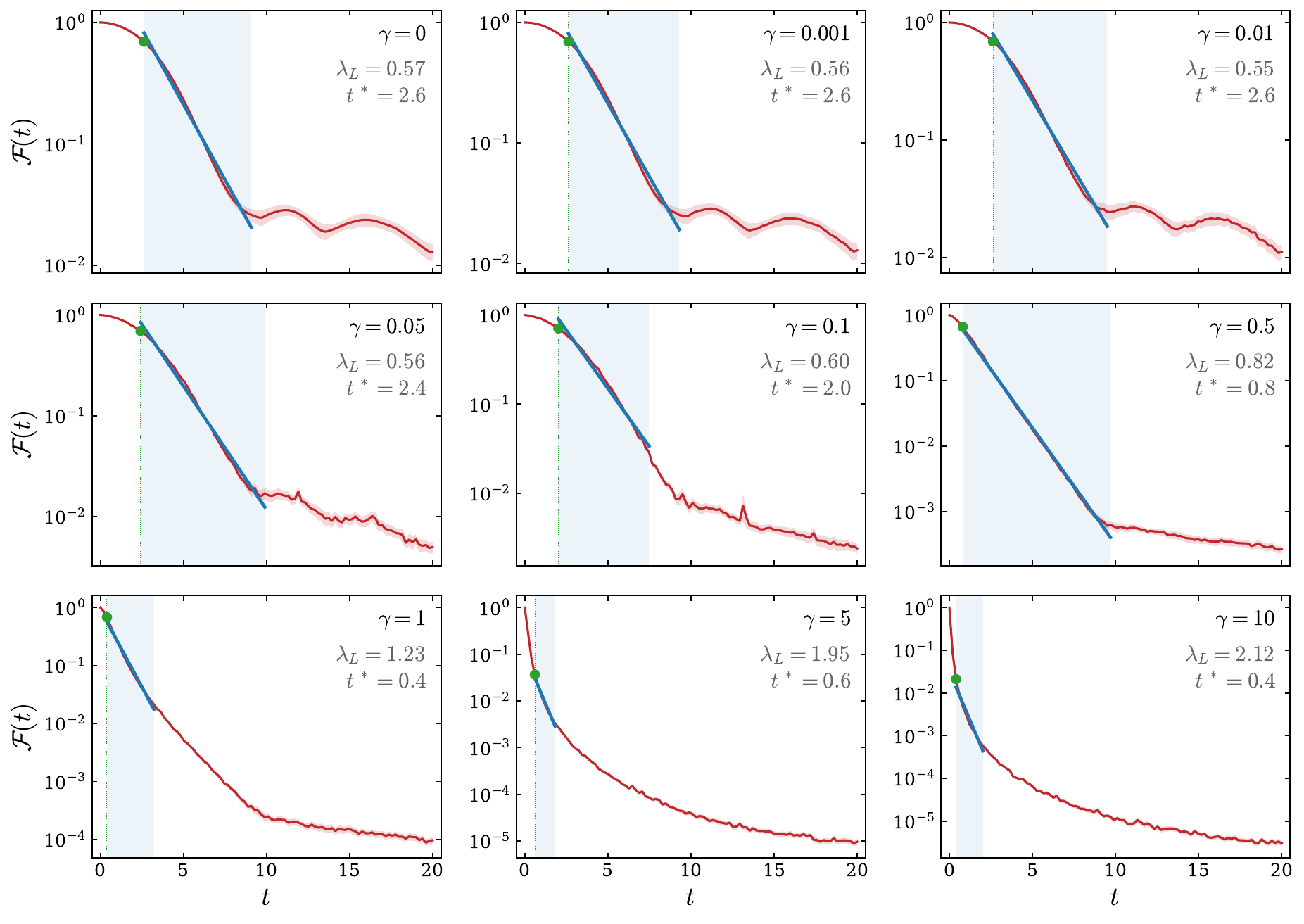}
    \caption{Averaged OTOC $\mathcal{F}(t)=|F(t)/F(0)|$ versus time for different monitoring rates (panels); red curves show the disorder-ensemble mean, averaged over 20 realizations, with shading giving the standard error. The intermediate regime
    is fit to $\mathcal{F}\sim e^{-\lambda_{\rm eff} t}$ (blue) over the adaptively
    chosen window (shaded), fixing the decay rate $\lambda_{\rm eff}$; the green marker
    locates the deviation time $t^{*}$ where the data peel away from the
    log-linear fit. With increasing $\gamma$, $\lambda_{\rm eff}$ grows and $t^{*}$
    contracts, while the late-time plateau rises then falls. Continuous
    monitoring steepens the OTOC-decay signature and shortens the window over
    which it holds.}\label{fig:otocgrid}
\end{figure}

 \begin{figure}
     \centering
     \includegraphics[width=0.75\columnwidth]{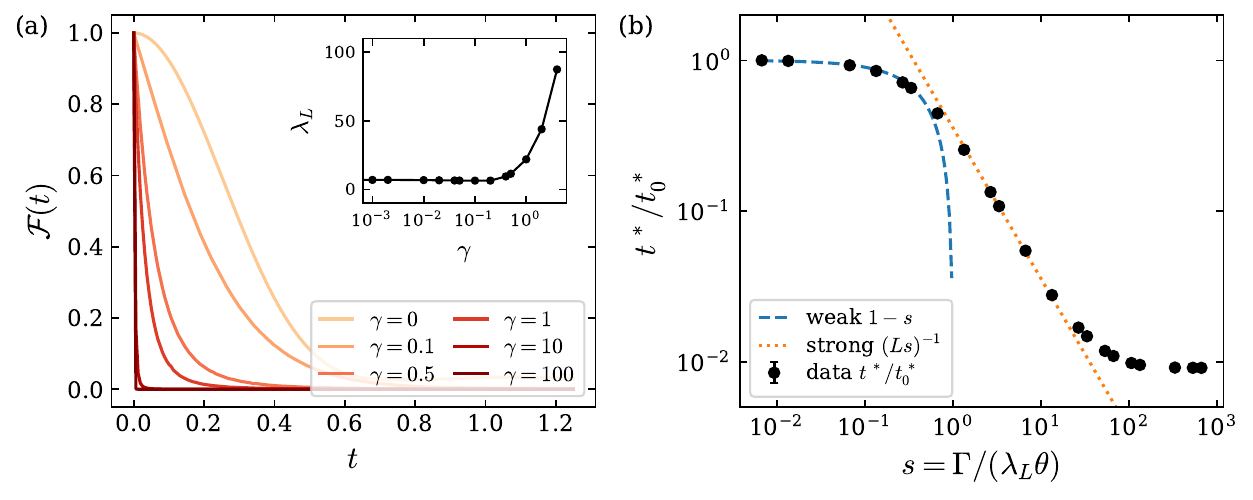}
     \caption{Monitored scrambling in GOE Hamiltonians. (a) Normalized OTOC
$\mathcal{F}(t)=F(t)/F(0)$ for a monitored GOE matrix of dimension $d=1024$, averaged over $100$ stochastic trajectories and $20$ RMT realizations, and for several monitoring strengths $\gamma$. Inset: fitted growth of the exponent $\lambda_{\rm eff}(\gamma)$. (b) Deviation time $t^*/t^*_0$ and versus $s=\Gamma/(\lambda_L\theta)$. The dashed and dotted lines show the weak-monitoring results, $1-s$ and the strong-monitoring result, $(Ls)^{-1}$, with $L=\lambda_L t^*_0$.}
     \label{suppfig:OTOC_GOE}
 \end{figure}

The competition between filtering and scrambling is also reflected in the deviation time $t^*$, defined as the earliest time at which $\mathcal F_\beta(t)$ departs from its fitted log-linear decay by a fixed fractional margin $\theta$. At weak monitoring the exponential regime survives, but the linear term of Eq.~\eqref{Eq:SM_OTOCdecay} brings the departure forward, so that
\begin{equation}\label{Eq:SM_tstarweak}
t^*(\gamma)\approx t^*_0\big(1-s\big),\qquad
s=\frac{\Gamma}{\lambda_L\,\theta} ,
\end{equation}
with $t^*_0$ the unmonitored value: $t^*$ decreases linearly with $\gamma$
at small monitoring strength.

For sufficiently strong monitoring, $\Gamma\gtrsim\lambda_L,$ the situation changes qualitatively. The energy filter suppresses the operator components before they have time to develop exponential growth. The intermediate scrambling window then disappears, and the decay of the OTOC is dominated by measurement backaction rather than chaotic operator spreading. In this regime,
\begin{equation}\label{Eq:SM_tstar}
\frac{t^*}{t^*_0}\simeq\frac{1}{L\,s},\qquad L=\lambda_Lt^*_0 ,
\end{equation}
until $t^*$ approaches the spacing of the time grid on which the fit is performed, beyond which the measured value saturates at a finite floor. The values extracted in Fig.~\ref{fig:otocgrid} are multiples of that spacing throughout, and the residual scatter among the largest $\gamma$ is of the same size. Strong monitoring therefore does not continue to advance the onset of scrambling indefinitely: it suppresses operator growth altogether, and beyond this point the OTOC no longer resolves the chaotic dynamics.

Overall, continuous measurement produces three regimes. At weak monitoring, the OTOC retains a clear chaotic growth window, while the measurement filter adds an extra decay and shortens the deviation time. At intermediate monitoring, this additional decay can strongly affect the exponent extracted from the OTOC. At strong monitoring, the filter suppresses operator growth before a scrambling regime can develop, leaving a $t^*\propto\gamma^{-1}$ collapse that ends in saturation at the grid resolution. The crossover occurs when the measurement rate becomes comparable to the intrinsic chaotic growth rate, $\Gamma\sim\lambda_L$, consistent with the scale at which the dissipative spectral diagnostics of the main text are extremal.

As shown in Fig.~\ref{suppfig:OTOC_GOE} our predictions of the scrambling analysis extend to GOE Hamiltonians. The normalized OTOC, fitted growth exponent, and deviation time exhibit the expected crossover from weak to strong monitoring, with the deviation-time scaling captured by the dimensionless parameter $s=\Gamma/(\lambda_L\theta)$. Our results thus clearly demonstrate that the enhancement of scrambling captured by the decay of the OTOC applies analogously to the broad class of chaotic systems with time-reversal symmetry as well.

\section{Liouvillian spectrum and dissipative spectral statistics under continuous monitoring}

As an additional open-system diagnostic, we analyze the spectrum of the stochastic generator $\mathcal L(dW_t)$ of the unnormalized trajectory state, retaining the noise term so that each trajectory carries its own Liouvillian and its own complex spectrum. In the energy eigenbasis $H\ket{n}=E_n\ket{n}$ all three terms act diagonally on the operator basis $\ket{n}\!\bra{m}$,
\begin{equation}
[H,[H,\ket{n}\!\bra{m}]]=(E_n-E_m)^2\,\ket{n}\!\bra{m},
\qquad
\{H,\ket{n}\!\bra{m}\}=(E_n+E_m)\,\ket{n}\!\bra{m},
\end{equation}
so the coherences are right eigenoperators and decouple into scalar stochastic equations $d\rho_{nm}=\Lambda_{nm}(dW_t)\,\rho_{nm}$ with
\begin{equation}
\Lambda_{nm}(dW_t)
=
\left[-i(E_n-E_m)-\gamma(E_n-E_m)^2\right]dt
+
\sqrt{2\gamma}\,(E_n+E_m)\,dW_t .
\end{equation}
As discussed above, It\^o integration gives
\begin{equation}
\rho_{nm}(t)
=
\exp\!\left[
-i(E_n-E_m)t
-\gamma(E_n-E_m)^2 t
-\gamma(E_n+E_m)^2 t
+\sqrt{2\gamma}\,(E_n+E_m)\,W_t
\right]\rho_{nm}(0),
\end{equation}
where the term $-\gamma(E_n+E_m)^2t$ is not present in the generator but is generated by the It\^o rule $dW_t^2=dt$ upon integrating the multiplicative noise. The full spectrum for given $\gamma$ and $W_t$ follows from a single diagonalization of the SYK spectrum $\{E_n\}$. The gap $E_n-E_m$ fixes the deterministic part of the complex eigenvalues, while the sum $E_n+E_m$ fixes the stochastic part, bypassing the $2^{N}\times2^{N}$ vectorization required for a non-commuting jump operator.

The single-time spectral object is
\begin{equation}\label{Eq:LambdaNM_SM}
\Lambda_{nm}^{(t,W_t)}
=
\frac{1}{t}\log\!\left[\frac{\rho_{nm}(t)}{\rho_{nm}(0)}\right]
=
-i(E_n-E_m)-\gamma(E_n-E_m)^2-\gamma(E_n+E_m)^2
+\sqrt{2\gamma}\,(E_n+E_m)\,\frac{W_t}{t},
\end{equation}
a set of $d^2$ complex numbers per realization, with $d$ the dimension of the parity sector retained. The drift is a one-dimensional locus parametrized by $(E_n,E_m)$. The noise shifts each point along the real axis by a Gaussian of variance $2\gamma(E_n+E_m)^2/t$. The eigenvalues thus form a two-dimensional cloud whose imaginary extent is set by the gap spectrum and whose real extent is set by $\gamma$, the energy sums, and $t$.

The complex spacing ratios probe the Liouvillian spectrum only through the two nearest neighbors of each eigenvalue. For measuring both spectral correlations of wider ranges and providing the open-system analog of dynamical signatures of dissipative quantum chaos, we study the dissipative spectral form factor (DSFF). The DSFF generalizes the ordinary SFF of Hamiltonian dynamics to the complex eigenvalues of the Liouvillian~\cite{Li2021,SenKumar2024}. Since the effective Liouvillian Eq.~\eqref{Eq:LambdaNM_SM} is diagonal, its full complex spectrum follows from a single Hamiltonian diagonalization, and the DSFF can be evaluated directly with the values $\{\Lambda_{nm}^{(t,W_t)}\}$. However, the corresponding time variable also needs to be considered in the full complex plane, parametrized as $T=|T|e^{i\varphi}$, due to the underlying non-hermiticity,
\begin{equation}\label{Eq:DSFFdef}
{\rm DSFF}(T)=\frac{1}{D}\left|\sum_{\alpha}
e^{\,i|T|\left(\cos\varphi\,\Re\Lambda_\alpha
+\sin\varphi\,\Im\Lambda_\alpha\right)}\right|^{2},
\end{equation}
where $\alpha=(n,m)$ runs over the $D=d^{2}-d$ elements of $\Lambda^{t,W_t}_{nm}$. At fixed $\varphi$, Eq.~\eqref{Eq:DSFFdef} is the ordinary form factor of the one-dimensional spectrum obtained by projecting the two-dimensional cloud along the direction $(\cos\varphi,\sin\varphi)$, so the
phase selects which combination of the gaps $E_n-E_m$ and the sums $E_n+E_m$ is
resolved.

We note that the DSFF factorizes exactly,
\begin{equation}\label{Eq:DSFFfact}
{\rm DSFF}(T)=\frac1D\Big|g_{\varphi}(T)\,g_{-\varphi}(T)-h(T)\Big|^2,
\end{equation}
with
\begin{equation}
g_{\pm\varphi}(T)=\sum_n e^{-i\left(\pm\tau_\pm E_n+\kappa E_n^2\right)},\qquad
h(T)=\sum_n e^{2i|T|\cos\varphi\,r_n},
\end{equation}
where $\tau_\pm=|T|\big(\sin\varphi\mp\sqrt{2\gamma}\cos\varphi\,W_t/t\big)$ is the coefficient of the phase linear in $E_n$, and
$\kappa=2\gamma|T|\cos\varphi$. In the simple limit of $\varphi=\pi/2$, $\kappa=0$ and $\tau_\pm\approx|T|$, recovering the ordinary SFF with the underlying Hamiltonian chaotic spectrum and with the modulus $|T|$, as the effective time parameter. Here, the noise term could also be neglected up to leading order for physical time scales near the dip time, as it scales down as $W_t/t\sim t^{-1/2}$. On the other hand, for $\varphi=0$, only the quadratic term survives, corresponding to the SFF with the Hamiltonian $H^2$, which slightly degrades the ramp by killing some of the spectral correlations with the squaring.

In the most interesting regime of intermediate $\varphi$, a similar energy filtering mechanism arises as in the stochastic SFF. The quadratic term now contributes to the phase of the DSFF rather than acting as a Gaussian cutoff. Instead, with increasing $\gamma$, it speeds up oscillations,  still acting as filtering out the spectral edges and extending the ramp. The slowest oscillating points of the exponents are obtained by the stationary points of  $\pm\tau_\pm E_n+\kappa E_n^2$,
\begin{equation}
E^*_\pm=\mp\frac{\tan\varphi}{4\gamma}+\frac{W_t}{2t\sqrt{2\gamma}},
\end{equation}
around which contributions to the DSFF are relevant in a window of $\kappa^{-1/2}=(2\gamma|T|\cos\varphi)^{-1/2}$. The window narrows with measurement strength exactly as the Gaussian filter $e^{-\gamma tE^2}$ of the stochastic SFF does, under the replacement $t\to|T|\cos\varphi$. The key difference in the two mechanisms is that here every term keeps unit modulus, with the suppression stemming from collective destructive interference, rather than individual suppression by the Gaussian term in the SFF. As the window scales also with $\cos\varphi$, for different phases the measurement strength acts differently, but overall the enhancement of the DSFF ramp behaves analogously to the SFF with the dip time matching the physical dip time up to constant phase factors, $|T_\mathrm{dip}|\sim t_\mathrm{dip}$. In particular, increasing the phase weakens the effects of $\gamma$ due to the scaling of the energy window $\sim(\gamma\cos\varphi)^{-1/2}$ and thus the optimal measurement strength $\gamma_\mathrm{opt}(\varphi)$ scales inversely with $\varphi$.

Figure~\ref{fig:DSFF} shows the averaged DSFF for the SYK model with $N=22$ (upper row) and the GOE with $d=1000$ (lower row) for  $\gamma=0,\,0.002,\,0.1,\,0.2,\,0.5$. As $F$ depends only on the spectrum $\{E_n\}$, the SYK and GOE results agree qualitatively. All curves start at the Liouville space dimension, $2^N$ and $d^2$, respectively, decrease once $|T|$ starts resolving the width of the spectrum, pass through the dip region either with or without a correlation hole, and saturate at the plateau ${\rm DSFF}=1$.
The plotted DSFFs agree well with the analytical predictions, as for both the SYK and RMT examples the maximally extended ramp appears for larger and larger values of $\gamma$ as $\varphi$ is increased. Additionally, Fig.~\ref{suppfig:CSR_GOE} shows the corresponding Liouvillian spectral analysis for GOE matrices, consistent with the SYK results shown in the End Matter. Overall, the Liouvillian analysis reveals a common qualitative structure in how monitoring reshapes spectral correlations across the complex spectrum, suggesting that these features are not specific to the underlying ensemble.

\begin{figure}
    \centering
    \includegraphics[width=1\columnwidth]{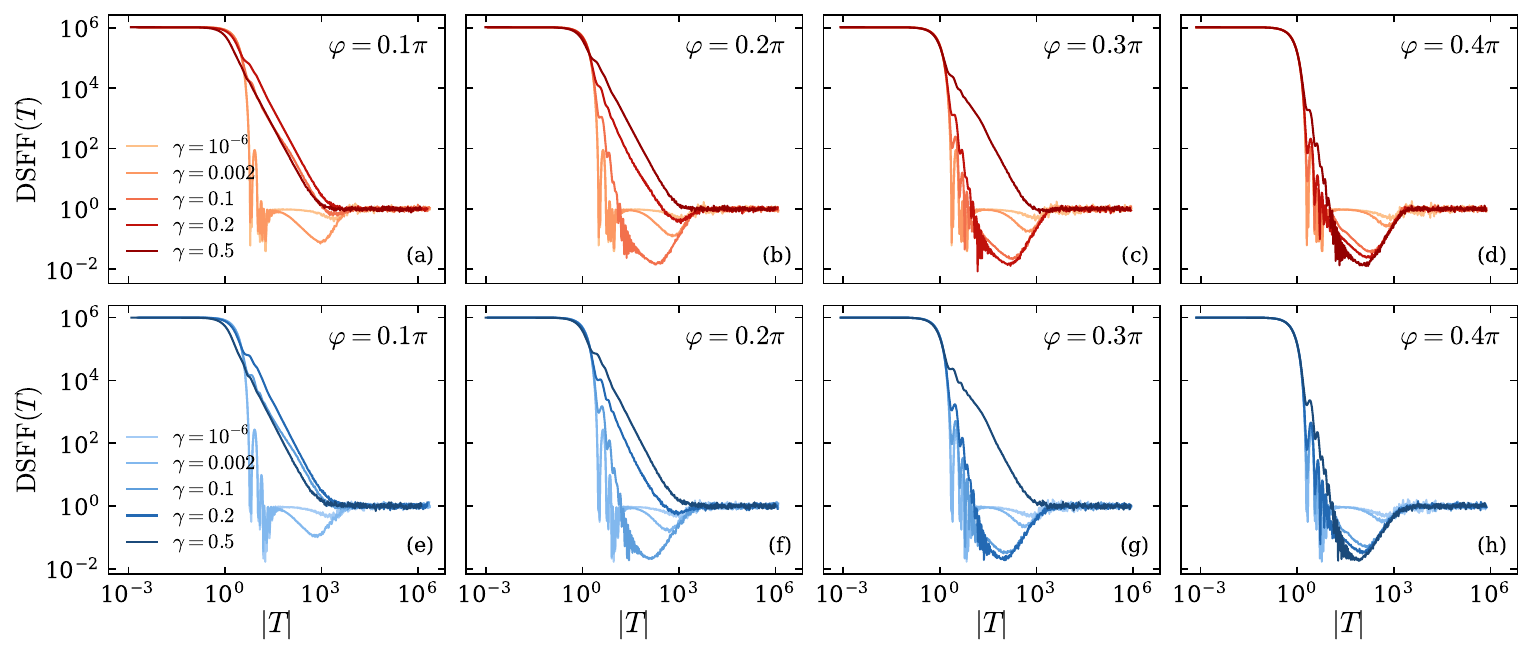}
    \caption{Dissipative spectral form factor built from the complex eigenvalues of the effective Wiener-noise-dependent Liouvillian plotted as a function of the complex time modulus for phases, $\varphi=0.1,(a),\,0.2,\,(b),\,0.3,\,(c),\,0.4,\,(d),\,$ and for $\gamma=0,\,0.002,\,0.1,\,0.2,\,0.5$. Upper row: The SYK model of $N=22$, for $\beta=0$ exhibiting the extension of the ramp as $\gamma$ increases, with the optimal measurement strength growing monotonically with $\varphi$ Lower row: Similar observations for the GOE RMT ensemble of Hilbert spaces dimension $N=1000$ and $\beta=0$ for the same values of $\gamma$ and complex time phases.}
    \label{fig:DSFF}
\end{figure}

\begin{figure}
    \centering
    \includegraphics[width=0.75\columnwidth]{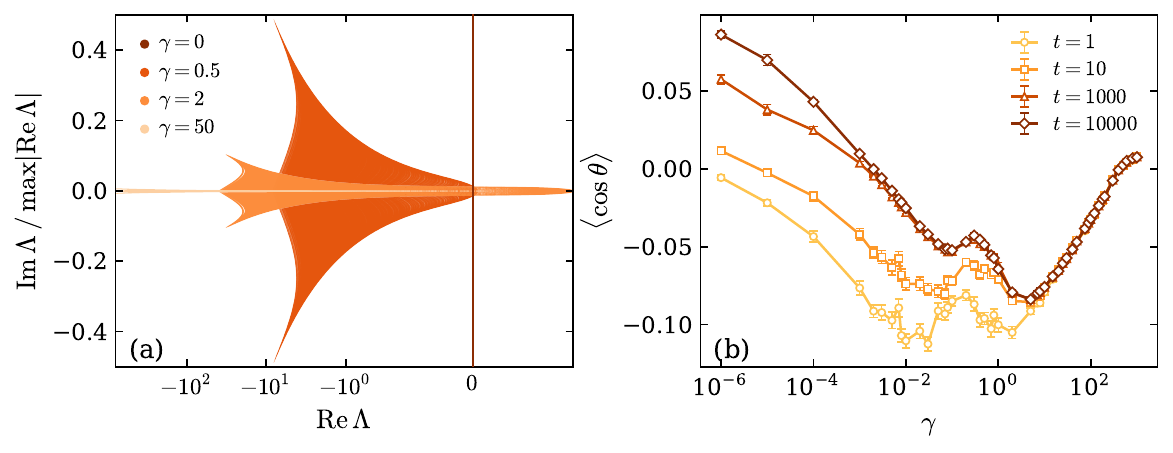}
    \caption{Dissipative Liouvillian spectrum of monitored GOE Hamiltonians ($d=1000$).
(a) Trajectory-resolved scatter of the Liouvillian eigenvalues $\Lambda_{nm}$
for different $\gamma$s. (b) Angular order parameter $\langle\cos\theta\rangle$, with $\theta$ the
argument of the complex spacing ratio, versus $\gamma$ for several
stroboscopic times $t$, averaged over $200$ realizations. GOE counterpart of Fig.~[4] of the End Matter.}
    \label{suppfig:CSR_GOE}
\end{figure}

\end{document}